# Nonemissive Intermediates: Photon-Yield Limits and Timing Advantages for Optogenetic Light Sources

Jingqi Chen,[*,†] Qiang Zhou,[*,‡,†] Yue Zhu,[†] and Xin Xu[*,¶,§]

† *Starlight Origin Technology Co., Ltd., Shenzhen, China*

‡ *Department of Mechanical Engineering, City University of Hong Kong, Hong Kong SAR, China*

¶ *Department of Materials Science and Engineering, National University of Singapore, Singapore*

§ *School of Materials Science and Engineering, Harbin Institute of Technology, Harbin, China*

E-mail: bestcjq@gmail.com; qzhou229-c@my.cityu.edu.hk; xin.xu@nus.edu.sg

**Abstract**

Maximizing photon yield need not minimize the excitation required for a brief optogenetic stimulus. We consider a reversible luminescent source in which a mechanically prepared excitation can reach the emitter directly or through one nonemissive intermediate. Dividing the mechanically prepared photon yield by the yield from direct optical excitation of the emitter gives the probability of successful arrival. When intermediate loss is positive, the same two yields predict the mean photoluminescence emission time given independently calibrated terminal rates and thermal weights. A separate optical decay trace can test this prediction. For a fixed budget of squared transfer couplings and allocation-independent terminal losses, we prove that either direct transfer or a pure relay attains the maximum total yield. Above a sharp threshold in the intermediate loss rate, direct transfer is globally optimal. In a spectrally weighted model of the light-gated channel Chrimson, a fixed relay in this regime emits 5.21% fewer photons per prepared excitation than direct transfer. Its earlier emission produces the same mean conductance over 20 ms with 16.54% less effective preparation and leaves lower residual conductance at 50 ms. Weak-drive response kernels show how early photons contribute to activation while leaving time for closure. Sustained stimulation can favor direct transfer again. Replaying calibrated emission waveforms with a common spectrum in voltage-clamped cells would test which source meets both conductance targets with fewer prepared excitations.

**TOC Graphic**

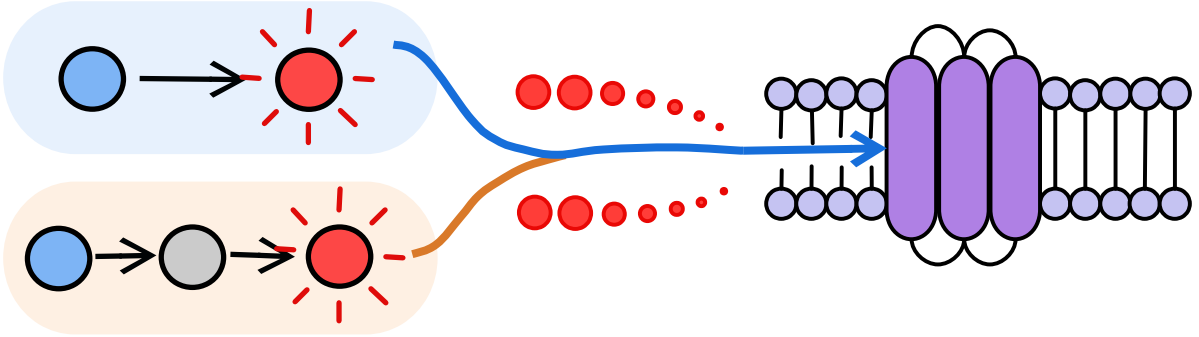

# 1 Introduction

A luminescent material must deliver photons with a suitable spectrum and time profile to drive a photosensitive receiver. In optogenetics, stimulation depends jointly on light generation, delivery, and molecular response. Miniaturized emitters and wireless implants provide local illumination.[1,2] Upconversion nanoparticles,[3–5] bioluminescent proteins,[6,7] and scintillators[8] offer further routes to generate activating light near the receiver. Mechanoluminescent and sonosensitized transducers extend these possibilities to ultrasound-triggered optical stimulation.[9–12] In luminescent transducers, an external stimulus prepares excitation that undergoes transfer, storage, and loss before emission. The output thus reflects both the amount of excitation prepared and its subsequent relaxation. Understanding this relaxation is essential because a higher photon yield need not make a source more effective for stimulation.

The receiver, in turn, determines which photon arrival times are useful. Channelrhodopsin engineering has produced bistable switches,[13] faster channels with reduced inactivation,[14,15] variants that respond efficiently at low light levels,[16] and red-shifted or highly sensitive tools.[17–19] Comparative studies reveal substantial differences in spectral sensitivity, photocurrent, and response kinetics.[20] Photocycle measurements and models trace these differences to transitions between conducting and nonconducting states.[21,22] Optical-stimulation models connect illumination to channel and neuronal dynamics.[23–25] Pulse shaping can therefore alter the light required for stimulation.[26,27] For a luminescent transducer, however, internal kinetics constrain the waveforms it can deliver. Even at equally effective wavelengths, photons that arrive in time for activation play a different role from those that prolong residual opening.

Materials design offers many ways to reshape these emission kinetics. Mechanoluminescent composites sustain repeated light generation under deformation,[28,29] and stress-responsive phosphors enable optical mapping of mechanical loads.[30] Near-infrared activators broaden the emission range,[31–33] while faulted nanocrystals bring mechanical excitation to smaller length scales.[34] Studies point to several microscopic mechanisms involving intrinsic defects,[35,36] heterojunction band offsets,[37] triboelectric excitation and interfacial charge transfer,[38,39] or tunneling-assisted trap release.[40] Time-resolved measurements also reveal frequency-dependent emission and evolving color.[41,42] These findings motivate separating excitation preparation from subsequent transfer to an emitting center. A defect or interface modification can change preparation efficiency, productive transfer, and loss simultaneously, making a ranking by intensity alone difficult to interpret.

The same competition between transfer and loss arises beyond mechanically excited phosphors. Resonant and exchange-mediated energy-transfer theories describe how excitation reaches emitting centers,[43,44] while nonradiative-transition theory accounts for competing routes to dissipation.[45] Triplet populations can feed delayed fluorescence through reverse intersystem crossing,[46,47] with intermediate electronic states facilitating spin conversion.[48] Charge separation and trapping can sustain persistent emission in organic donor–acceptor materials[49–51] and inorganic phosphors.[52,53] State-resolved studies of upconversion nanocrystals similarly distinguish productive transfer from quenching.[54] We consider the kinetic features these systems share when their population dynamics admit a closed, single-excitation Markov description with time-independent rates and internal detailed balance.

Yield and decay time also depend on processes at the emitting center itself. Single-molecule nanoantenna experiments distinguish excitation enhancement, radiative acceleration, and metal-induced quenching,[55–57] while plasmonic cavities can achieve both rapid spontaneous emission and high radiative efficiency.[58] Internal transfer introduces a dependence on preparation: an excitation prepared outside the emitter must first reach it, whereas one prepared in the emitter can radiate immediately. An intermediate may therefore aid entry while also opening routes for excursions and loss after arrival. Preparation-resolved yields and emission times capture these competing roles.

Established kinetic methods offer a framework for this analysis. Compartmental photophysics identifies when fluorescence measurements can determine transfer rates and spectral parameters.[59–61] Analyses of projected relaxation and single-molecule trajectories further show how underlying states can remain hidden in measured signals.[62,63] First-passage theory isolates access to a target,[64,65] and its connection to reversible relaxation provides a basis for separating arrival from subsequent motion.[66,67] Building on these methods, we ask which combinations of yield and emission timing a reversible source can produce, and when redistributing a fixed transfer resource can improve its output.

We follow a single excitation through the reversible network in Section 2 (Figure 1). First arrival at the emitter separates the probability of reaching it from the probability of emitting once there. In Section 3, this separation and detailed balance show how the yields after mechanical and optical preparation determine the mean photoluminescence emission time for positive intermediate loss at fixed terminal rates and thermal weights (Figure 2). Section 4 then asks whether reallocating a fixed squared-coupling budget can increase total yield. With allocation-independent terminal losses, direct transfer or a pure relay always attains the optimum (Figure 3).

We test the value of photon timing above the intermediate-loss threshold, where direct transfer already maximizes total yield. Section 5 compares that source with a lower-yield relay in a spectrally weighted Chrimson benchmark. We first compare receiver responses at equal preparation (Figure 4), then the preparation each source needs for the same activation and closure targets (Figure 5). Section 6 asks when the ranking reverses as the response window changes or stimulation is sustained (Figure 6).

## 2 From a prepared excitation to a receiver task

We follow a single excitation prepared in the mechanical configuration $C$ or the emissive state $E$. Transfer is reversible along a direct route ($C \leftrightarrow E$) and an indirect route through the nonemissive intermediate $I$ ($C \leftrightarrow I \leftrightarrow E$), as shown in Figure 1(a). The excitation ends at the absorbing ground state $G$ through monitored emission from $E$ or competing losses from $C$, $I$, and $E$.

Defect-assisted transfer in inorganic mechanoluminescent materials motivates the intermediate configuration.[68–71] Each state represents a configuration or a group with closed population dynamics. This description follows either from exact grouping or, approximately, from rapid intragroup thermalization. Bath conditions are fixed, and population-relevant coherence is negligible. Emission outside the monitored channel counts as terminal loss only if it neither re-excites the source nor drives the receiver. The grouping and material-specific conditions are detailed in SI1.2.

$ZnS:Mn^{2+}$ provides a qualitative material connection. Single-particle measurements locate mechanoluminescence (ML) in regions with weak Mn photoluminescence (PL). The proposed excitation scheme includes local carrier recombination and a route through a Mn-associated bound exciton.[72] In a local single-excitation description, $C$ can represent the prepared carrier pair and $I$ a bound-exciton configuration with a separately resolved population. The emitter $E$ is the $Mn^{2+}$ ${}^4T_1$ manifold, and $G$ denotes termination with Mn returned to ${}^6A_1$. These assignments draw on the proposed Mn excitation routes. The benchmark specifies the reversible links and their rates independently of the material data. Sample-specific tests are discussed in SI1.4.

Let $a$ and $b$ denote terminal loss rates from $C$ and $I$. State $E$ emits monitored photons at rate $r$ or terminates at rate $n$, with $\kappa = r + n$. In the numerical realization, $a$, $b$, and $n$ are nonradiative rates. Using the transfer rates in Figure 1(a) and the column-population vector $\boldsymbol{p} = (p_C, p_I, p_E)^\mathsf{T}$, we write

$$\dot{\boldsymbol{p}} = L\boldsymbol{p}, \qquad L = \begin{pmatrix} -(a+c+u) & v & d \\ u & -(b+v+w) & z \\ c & w & -(\kappa+d+z) \end{pmatrix}. \tag{1}$$

During relaxation, $G$ is absorbing and $p_G = 1 - p_C - p_I - p_E$. Internal detailed balance requires

$$d = \chi_C c, \qquad v = \frac{\chi_C u}{\chi_I}, \qquad z = \chi_I w.$$

At bath temperature $T_b$, the thermal weights are

$$\chi_i = \frac{Z_i}{Z_E} = \exp\left[-\frac{F_i^{\text{th}} - F_E^{\text{th}}}{k_B T_b}\right], \qquad F_i^{\text{th}} = -k_B T_b \ln Z_i. \tag{2}$$

Each $Z_i$ includes the group's internal multiplicity. The weights give equilibrium population ratios for the internal network with termination disabled. An equilibrated threefold triplet $I$ and singlet $E$ with otherwise equal partition factors give $\chi_I = 3\exp[-(\epsilon_I - \epsilon_E)/(k_B T_b)]$. The

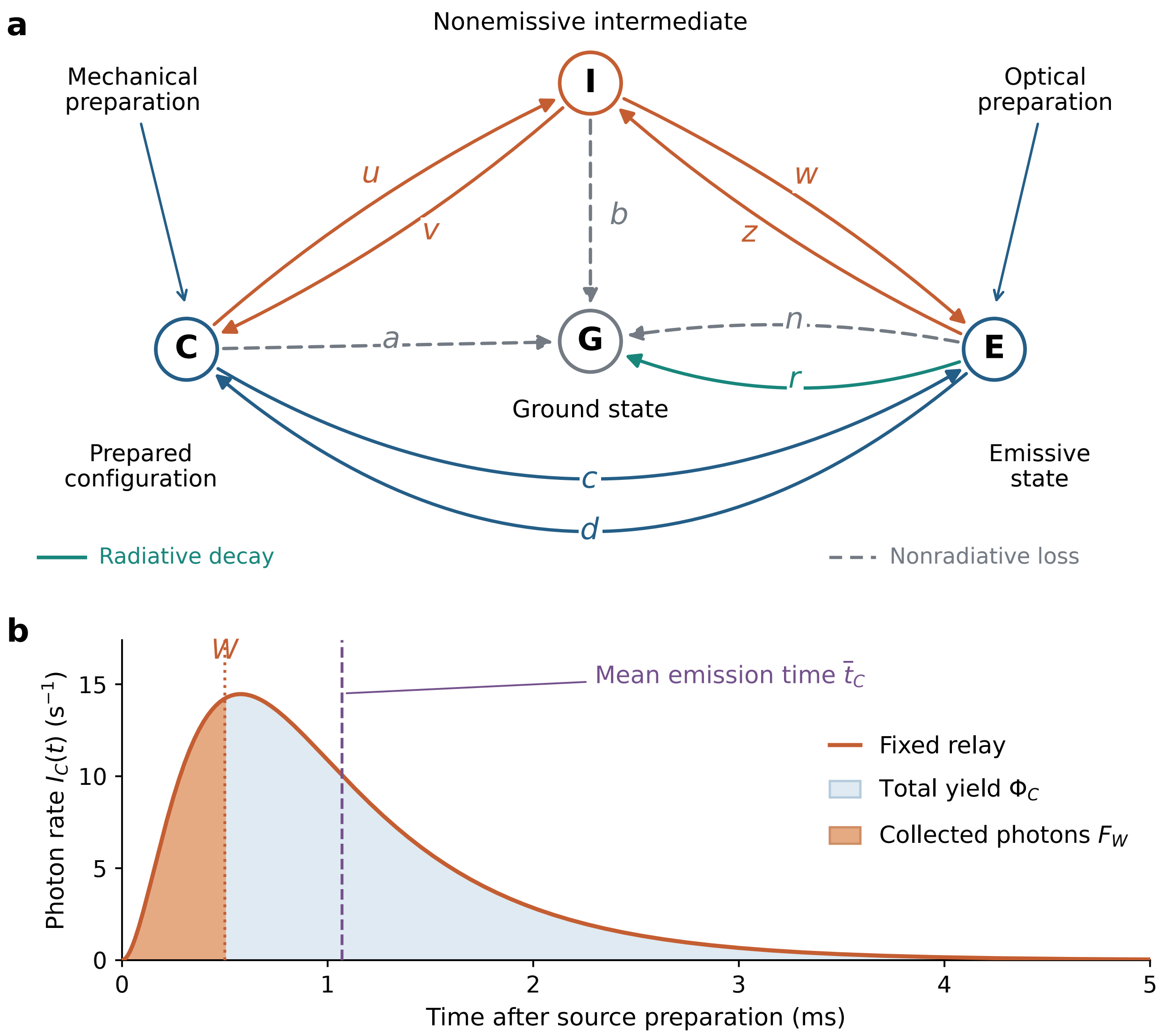

Figure 1: An excitation prepared mechanically in $C$ must reach $E$ to emit. (a) Optical preparation starts in $E$. Reversible transfer connects $C$ and $E$ directly $(c, d)$ or through the nonemissive intermediate $I$ ($u, v$ on $C \leftrightarrow I$ and $w, z$ on $I \leftrightarrow E$). Emission ($r$, green) and nonradiative loss ($a, b, n$, dashed gray) end at $G$. (b) For the fixed relay, the orange area under the photon rate $I_C(t)$ gives the yield $F_W$ by $W = 0.5$ ms. The full area, including the tail beyond 5 ms, gives the total yield $\Phi_C$. The dashed purple line marks the conditional mean emission time $\bar{t}_C$.

free-energy gap therefore contains $-k_BT_b \ln 3$, while the $\epsilon_i$ denote potential minima.[73]

The numerical realization uses one nondegenerate displaced harmonic surface per state, with minimum $\epsilon_i$ and common internal partition factors. The thermal Marcus–Jortner rates are quadratic in the electronic couplings and obey internal detailed balance. The surfaces, sideband weights, and full rate expression are given in SI1.1.[74–76]

For general transfer rates, first-arrival factorization requires Markov dynamics and emission from a single state. The yield–time closure also uses internal detailed balance and the single-intermediate topology. The allocation theorem further requires a fixed squared-coupling budget and allocation-independent terminal losses.

For a unit preparation in $i \in \{C, E\}$, set $N = (-L)^{-1}$. The photon rate, total monitored yield, and conditional mean emission time are

$$\begin{aligned} I_i(t) &= r\boldsymbol{e}_E^{\mathsf{T}} e^{Lt} \boldsymbol{e}_i, \\ \Phi_i &= \int_0^\infty I_i(t)\, dt = rN_{Ei}, \\ \bar{t}_i &= \frac{\int_0^\infty tI_i(t)\, dt}{\Phi_i} = \frac{(N^2)_{Ei}}{N_{Ei}}. \end{aligned} \tag{3}$$

Here $\boldsymbol{e}_i$ denotes unit population in state $i$. The inverse acts on the accessible transient subspace, and the mean is defined for $\Phi_i > 0$. Figure 1(b) distinguishes the full area $\Phi_C$, the area collected before $W$, $F_W = \int_0^W I_C(t)\, dt$, and the conditional mean $\bar{t}_C$.

We call $\Phi_C$ the mechanically prepared luminescence yield and $\Phi_E$ the PL yield. For $N_C$ independent preparations, the mean photon count is $N_C\Phi_C$. Relating $N_C$ to applied mechanical work requires an independent calibration of preparation efficiency. Preferential loss of late-emitting trajectories can shorten the conditional mean, so yield and time must be interpreted together.

To compare transfer routes under a common resource constraint, we fix the total squared coupling across the three internal links:

$$|V_{CE}|^2 + |V_{CI}|^2 + |V_{IE}|^2 = V_B^2,$$
$$(f_{CE}, f_{CI}, f_{IE}) = (1-\eta, \eta s, \eta(1-s)), \tag{4}$$
$$c = c_0(1-\eta), \qquad u = u_B \eta s, \qquad w = w_B \eta(1-s).$$

Here $f_{ij} = |V_{ij}|^2/V_B^2$ and $0 \le \eta, s \le 1$. Each forward rate is linear in the fraction allocated to its link. Assigning the full budget to a single link gives $c_0$ for $C \to E$, $u_B$ for $C \to I$, or $w_B$ for $I \to E$. A fraction $\eta$ goes to the indirect route; $s$ splits it between $C \leftrightarrow I$ and $I \leftrightarrow E$. Reverse rates follow from detailed balance; in particular, $v = v_B \eta s$ with $v_B = \chi_C u_B/\chi_I$.

Redistribution holds the state energies, geometry, temperature, and terminal rates fixed. Each link's energy-dependent and vibrational rate factor therefore remains unchanged, and terminal channels lie outside the budget. A material modification that changes energies or losses requires updated coefficients and a new threshold. The direct reference has $\eta = 0$; a pure relay has $\eta = 1$ and a specified split $s$.

We measure state energies relative to the emitter, $\Delta_i = \epsilon_i - \epsilon_E$ for $i = C, I$. The central example uses $T_b = 300$ K and $\Delta_I = -20$ meV. For the receiver benchmark, we shift the excited-state energies by a common amount relative to $G$ and adjust the terminal couplings and radiative dipole. This preserves the kinetics while giving a calculated photon-spectrum peak at 594 nm. Section 5 weights the full broadband emission spectrum by the receiver's spectral sensitivity; the spectral construction is specified in SI1.3.

The relay allocation is fixed at $\eta = 1$ and $s \simeq 0.4942$, with its $C$-prepared emission curve shown in Figure 1(b). This split was selected for early photon collection and is held fixed across receiver response windows. Figure 2 varies energy and loss around this construction, while Figure 3 explores the full allocation simplex. Source parameters and receiver normalization are specified in SI1 and SI2.

# 3 Preparation resolves emission regimes and constrains their observables

Mechanical and optical preparation differ in how an excitation enters the emitter. An excitation prepared in $C$ must first reach $E$; one prepared in $E$ starts there. After the first arrival at $E$, both preparations share the same emission dynamics. Separating entry from this common response gives exact relations between their waveforms, yields, and mean times. The same intermediate can therefore enhance mechanically prepared emission, quench PL, and shift different measures of emission time in opposite directions.

## First arrival separates access from emission

Let $T_E$ and $T_G$ be the first arrival times at $E$ and $G$. For preparation in $i = C, I$, let $f_i(t)$ be the density of first reaching $E$ before termination. This density is defective: some excitations terminate before reaching $E$, so its integral can be less than one. The entry probability and unnormalized first moment are

$$\begin{aligned} q_i &= \int_0^\infty f_i(t)\,dt = \Pr_i(T_E < T_G), \\ m_i &= \int_0^\infty t f_i(t)\,dt. \end{aligned}$$

Every photon emitted after $C$ preparation is preceded by a successful first arrival at $E$. The Markov property then gives

$$I_C(t) = \int_0^t f_C(s) I_E(t-s)\,ds. \tag{5}$$

Integration and the first moment reduce this waveform relation to

$$\boxed{\Phi_C = q_C \Phi_E, \qquad \bar{t}_C = \bar{t}_E + \frac{m_C}{q_C}}. \tag{6}$$

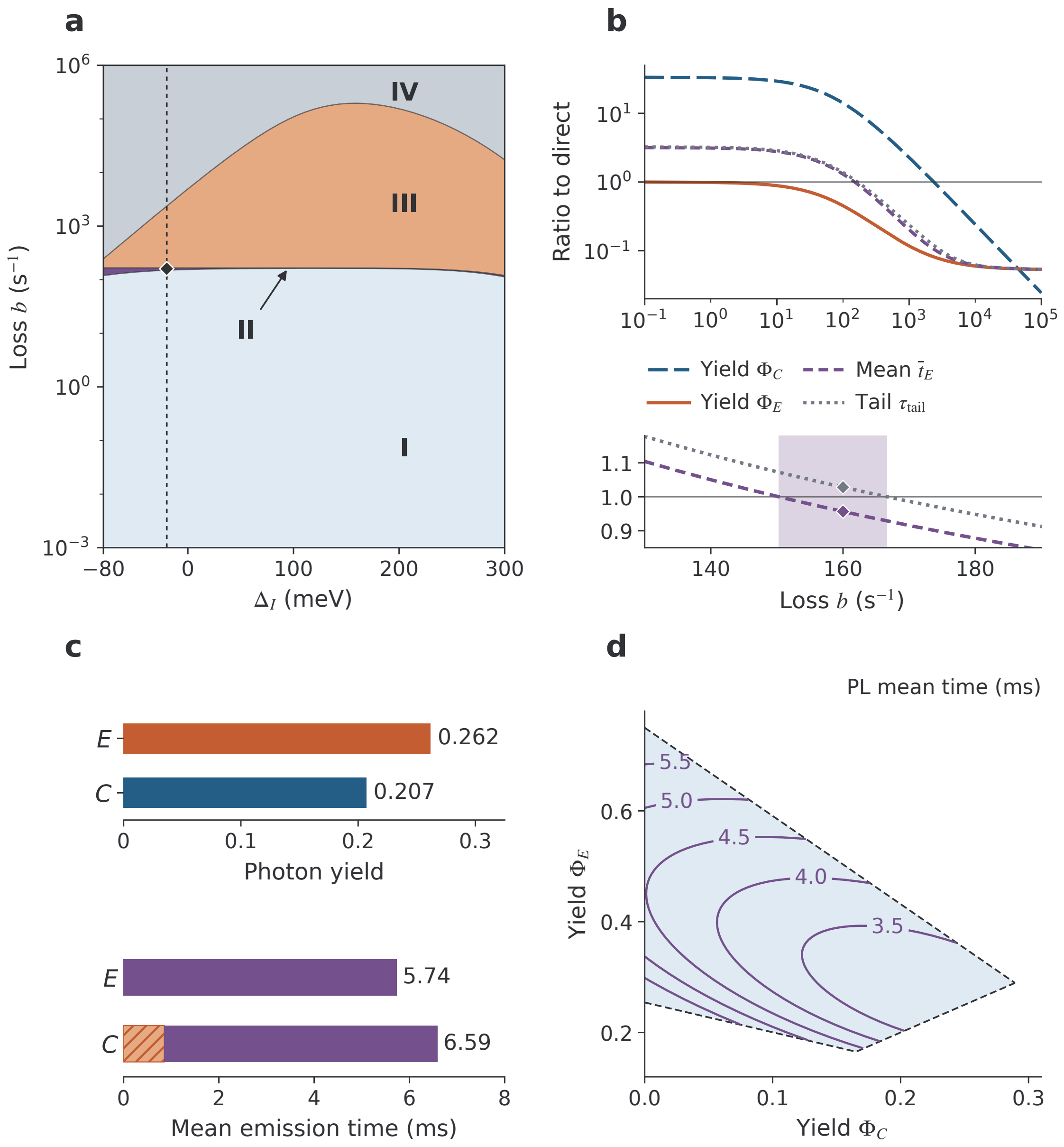

Figure 2: A relay can raise the mechanically prepared yield $\Phi_C$ while lowering the photoluminescence (PL) yield $\Phi_E$. (a) The fixed pure relay ($s = 0.49417$) shows this combination in regions I–III relative to direct transfer at the same budget. Both yields fall in IV. The PL mean emission time and asymptotic decay time both lengthen in I and shorten in III and IV. (b) At $\Delta_I = -20$ meV, shading highlights region II, where the mean shortens and the tail lengthens. (c) At $b = 160\,\mathrm{s}^{-1}$ (diamonds in (a,b)), the yield ratio $\Phi_C/\Phi_E = 0.791$ gives the probability of reaching $E$ from $C$. Successful entry takes 0.854 ms on average (hatched difference in means). (d) For $b > 0$ and fixed terminal rates and thermal weights, the two yields determine the PL mean (contours in ms). Dashed edges mark the probability bounds in Eq. (12). Here $(\Delta_C, \Delta_I) = (+20, -20)$ meV and $b = 150\,\mathrm{s}^{-1}$. Panels (a–c) use $\Delta_C \simeq 306.2$ meV. Other source parameters are given in SI1.

The mean-time relation is defined for $\Phi_C > 0$. The yield ratio gives the entry probability; the mean-time difference gives the mean duration of successful entry. Detailed balance enters only in the constraints below.

Figure 2(c) displays these quantities for two unit-normalized preparations. The hatched segment adds the conditional entry time to the common post-entry mean.

First-step conditioning gives both entry statistics from the same two-state operator:

$$\begin{aligned} H &= \begin{pmatrix} a+c+u & -u \\ -v & b+v+w \end{pmatrix}, \\ H\boldsymbol{q} &= \begin{pmatrix} c \\ w \end{pmatrix}, \qquad H\boldsymbol{m} = \boldsymbol{q}. \end{aligned} \tag{7}$$

The Laplace transforms, $\widehat{f}_i(\zeta) = \int_0^\infty e^{-\zeta t} f_i(t)\,dt$, satisfy

$$\widehat{\boldsymbol{f}}(\zeta) = (H + \zeta \mathbf{I}_2)^{-1}(c, w)^\mathsf{T},$$

where $\mathbf{I}_2$ is the identity on the $(C, I)$ subspace. After entry, an excitation can emit, terminate nonradiatively, or leave $E$ and return. Summing repeated returns gives the renewal relation[67]

$$\begin{aligned} \widehat{I}_E(\zeta) &= \frac{r}{\zeta + \kappa + d + z - d\widehat{f}_C(\zeta) - z\widehat{f}_I(\zeta)}, \\ \widehat{I}_C(\zeta) &= \widehat{f}_C(\zeta)\widehat{I}_E(\zeta). \end{aligned} \tag{8}$$

At $\zeta = 0$, the transform and its first derivative give

$$\begin{aligned} \Phi_E &= \frac{r}{\Lambda_E}, \qquad \Lambda_E = \kappa + d(1 - q_C) + z(1 - q_I), \\ \bar{t}_E &= \frac{M}{\Lambda_E}, \qquad M = 1 + dm_C + zm_I. \end{aligned} \tag{9}$$

Thus $q_C$ measures access, $\Lambda_E$ includes loss through failed excursions, and $M$ accounts for the duration of excursions that return. Since $q_C \le 1$ and $\Lambda_E \ge \kappa$,

$$0 \le \Phi_C \le \Phi_E \le \frac{r}{\kappa}.$$

## Entry, loss, and emission regimes

The inequality $\Phi_C \le \Phi_E$ compares two preparations of the same source, as in Figure 2(c). To assess the relay, we now compare the two sources at the same preparation, as in Figure 2(b). Taking direct transfer as the reference (subscript 0), define

$$R_{\text{entry}} = \frac{q_C}{q_{C0}}, \qquad R_{\text{loss}} = \frac{\Lambda_E}{\Lambda_{E0}}, \qquad R_{\text{memory}} = \frac{M}{M_0}.$$

The observable ratios become

$$\frac{\Phi_C}{\Phi_{C0}} = \frac{R_{\text{entry}}}{R_{\text{loss}}}, \qquad \frac{\Phi_E}{\Phi_{E0}} = \frac{1}{R_{\text{loss}}}, \qquad \frac{\bar{t}_E}{\bar{t}_{E0}} = \frac{R_{\text{memory}}}{R_{\text{loss}}}. \tag{10}$$

Enhanced mechanically prepared emission with quenched PL requires $1 < R_{\text{loss}} < R_{\text{entry}}$: the entry gain must outweigh the added loss after arrival. A longer PL mean requires the separate condition $R_{\text{memory}} > R_{\text{loss}}$. Improved entry alone therefore does not set the direction of change in the PL mean.

Figure 2(a) expresses these regimes in the physical parameters $\Delta_I$ and $b$. At the fixed pure-relay allocation, the domain $\Delta_I \in [-80, 300]$ meV and $b \in [10^{-3}, 10^{6}]\,\text{s}^{-1}$ contains four regimes. Regions I–III enhance the mechanically prepared yield while quenching PL. Their PL timing differs: both the mean and asymptotic tail are longer in I, only the tail is longer in II, and both are shorter in III. Region IV also loses the mechanically prepared yield advantage.

The loss scan in Figure 2(b) resolves the narrow region II. At $b = 160\,\text{s}^{-1}$, the mechanically prepared yield rises to 10.47 times the direct value, while the PL yield falls to 0.349 of its direct value. The PL mean is 4.34% shorter, but the asymptotic tail is 2.82% longer. The later source–receiver comparisons use $b = 2500\,\text{s}^{-1}$, beyond the mechanically prepared yield reversal in this scan. The horizontal energy scan holds the actual loss $b$ fixed by adjusting $V_{IG}$ within the microscopic realization.

## An observable yield–time constraint

Detailed balance turns the entry statistics into an observable constraint. Reciprocity gives $N_{iE} = \chi_i N_{Ei}$, while first arrival gives $N_{Ei} = q_i N_{EE}$. Substituting these identities into $(N^2)_{EE} = \sum_j N_{Ej} N_{jE}$ yields

$$\begin{aligned} M &= 1 + \chi_C q_C^2 + \chi_I q_I^2, \\ \bar{t}_E &= \frac{\Phi_E}{r}\left(1 + \chi_C q_C^2 + \chi_I q_I^2\right). \end{aligned} \tag{11}$$

The time factor now contains only entry probabilities and thermal weights, including internal degeneracy.

Set $P = \Phi_E$, $Q = \Phi_C$, and $\tau = \bar{t}_E$. Integrated probability conservation after $E$ preparation, together with $q_C = Q/P$, gives

$$1 = aN_{CE} + bN_{IE} + \kappa N_{EE},$$

$$\mathcal{R} = r - \kappa P - a\chi_C Q = b\chi_I P q_I.$$

Eliminating $q_I$ from Eq. (11) gives, for $b > 0$,

$$\begin{aligned} &\boxed{r\tau P = P^2 + \chi_C Q^2 + \frac{\mathcal{R}^2}{b^2 \chi_I}}, \\ &0 \le Q \le P, \qquad 0 \le \mathcal{R} \le b\chi_I P. \end{aligned} \tag{12}$$

At fixed terminal rates and thermal weights, the two yields fix the PL mean. No individual internal transfer rate appears in this surface relation.

Figure 2(d) shows the surface as contours of constant PL mean in the two-yield plane. To display the dependence on both yields, this separate analytical example uses $\Delta_C = +20$ meV;

the remaining parameters are specified in the caption. Shading marks the domain allowed by the probability constraints. A specified microscopic coupling budget further restricts its accessible subset.

If $b$ is unknown, $0 \leq q_I \leq 1$ still gives

$$P^2 + \chi_C Q^2 \leq r\tau P \leq (1 + \chi_I)P^2 + \chi_C Q^2. \tag{13}$$

Conversely, the three observables can determine the loss rate within the single-intermediate model:

$$b = \frac{r - \kappa P - a\chi_C Q}{\sqrt{\chi_I \left(r\tau P - P^2 - \chi_C Q^2\right)}}. \tag{14}$$

This inference requires known $a$, $r$, $\kappa$, and thermal weights, absolute preparation normalization, and positive numerator and denominator. Nearly vanishing terms make the estimate ill-conditioned. An independent test of the closure relation requires separately calibrated inputs and a mean-time measurement reserved for validation. Normalization and calibration procedures are given in SI1.4.

For several nonemissive states $I_\alpha$, the corresponding relations are

$$\mathcal{R} = P \sum_\alpha b_\alpha \chi_\alpha q_\alpha,$$

$$r\tau P - P^2 - \chi_C Q^2 = P^2 \sum_\alpha \chi_\alpha q_\alpha^2.$$

Cauchy–Schwarz then gives

$$r\tau P - P^2 - \chi_C Q^2 \geq \frac{\mathcal{R}^2}{\sum_\alpha b_\alpha^2 \chi_\alpha}, \tag{15}$$

provided the denominator is positive. The single-intermediate equality becomes a lower bound for a larger reversible network. Additional states can therefore move the observables off the single-intermediate surface while preserving reversibility.

## Mean time, asymptotic tail, and controlled loss

Reversibility also constrains the PL mean and asymptotic tail. Symmetrizing the generator gives

$$I_E(t) = r \sum_j A_j e^{-\lambda_j t}, \qquad A_j \geq 0, \qquad \sum_j A_j = 1,$$

with $\lambda_j > 0$ on the accessible transient subspace. The mean is a weighted average of modal decay times; the tail is set by the slowest mode with nonzero PL weight. Hence

$$\frac{P}{r} \leq \tau \leq \tau_{E,\mathrm{tail}} = \frac{1}{\min_{j:A_j>0} \lambda_j}. \tag{16}$$

The lower bound follows from Cauchy–Schwarz, and the upper bound from nonnegative modal weights. Both the weights and decay rates change across sources, allowing the mean and tail to move in opposite directions relative to a reference, as in Figure 2(b). A fit over a finite time window can include several decay modes.

Varying $b$ while holding all other rates fixed gives

$$\partial_b N = -N \boldsymbol{e}_I \boldsymbol{e}_I^{\mathsf{T}} N,$$

and hence

$$\begin{aligned} \partial_b P &= -\frac{\chi_I}{r}(q_I P)^2, \\ \tau &= \frac{P}{r}\left[1 + \chi_C \left(\frac{Q}{P}\right)^2\right] - \partial_b \ln P. \end{aligned} \tag{17}$$

The logarithmic yield response to $b$ thus measures the intermediate's contribution to the PL mean. At fixed transfer rates and other terminal rates, yields and hitting probabilities are nonincreasing in $b$. Equation (11) gives the same monotonicity for the PL mean. The slowest decay rate with nonzero PL weight is nondecreasing, so the PL tail time is also nonincreasing. During a pure loss scan, these observables can cross a fixed reference only in the decreasing direction.

# 4 A fixed coupling budget imposes a global yield limit

In Section 3, increasing the intermediate loss $b$ eventually erases the fixed relay's yield advantage over direct transfer. At the same loss, can a different allocation of the coupling budget restore that advantage? For the reversible source with $a, c_0, r > 0$ and $b \geq 0$, hold the squared-coupling budget of Eq. (4) and the terminal rates fixed. The total photon yield per $C$ preparation then obeys

$$\boxed{\begin{aligned} &\Phi_C(\eta, s) \leq \max\{\Phi_{C0}, \Phi_C(1, s)\}, \\ &\max_{\eta,s} \Phi_C = \max\{\Phi_{C0}, \max_s \Phi_C(1, s)\}. \end{aligned}} \tag{18}$$

At each split $s$, the better endpoint bounds every mixture. The global optimum therefore follows by comparing direct transfer with the best pure relay [Figure 3(a,b)].

## Why a small shift toward the relay costs yield

Fix the split $s$ between the two relay links and write

$$p = u_B s, \qquad h = w_B(1 - s), \qquad \nu = v_B s.$$

As the relay fraction $\eta$ increases from zero to one, the direct rate falls as $c = c_0(1 - \eta)$ and the relay rates grow as $u = p\eta$, $v = \nu\eta$, and $w = h\eta$.

Integrating the population equations over the full decay lets us eliminate $I$ exactly. With $\mathcal{B} = b + v + w > 0$, the effective rates are

$$\begin{aligned} c_{\text{eff}} &= c + \frac{uw}{\mathcal{B}}, \quad a_{\text{eff}} = a + \frac{ub}{\mathcal{B}}, \\ d_{\text{eff}} &= \chi_C c_{\text{eff}}, \quad \kappa_{\text{eff}} = \kappa + \frac{\chi_I wb}{\mathcal{B}}. \end{aligned}$$

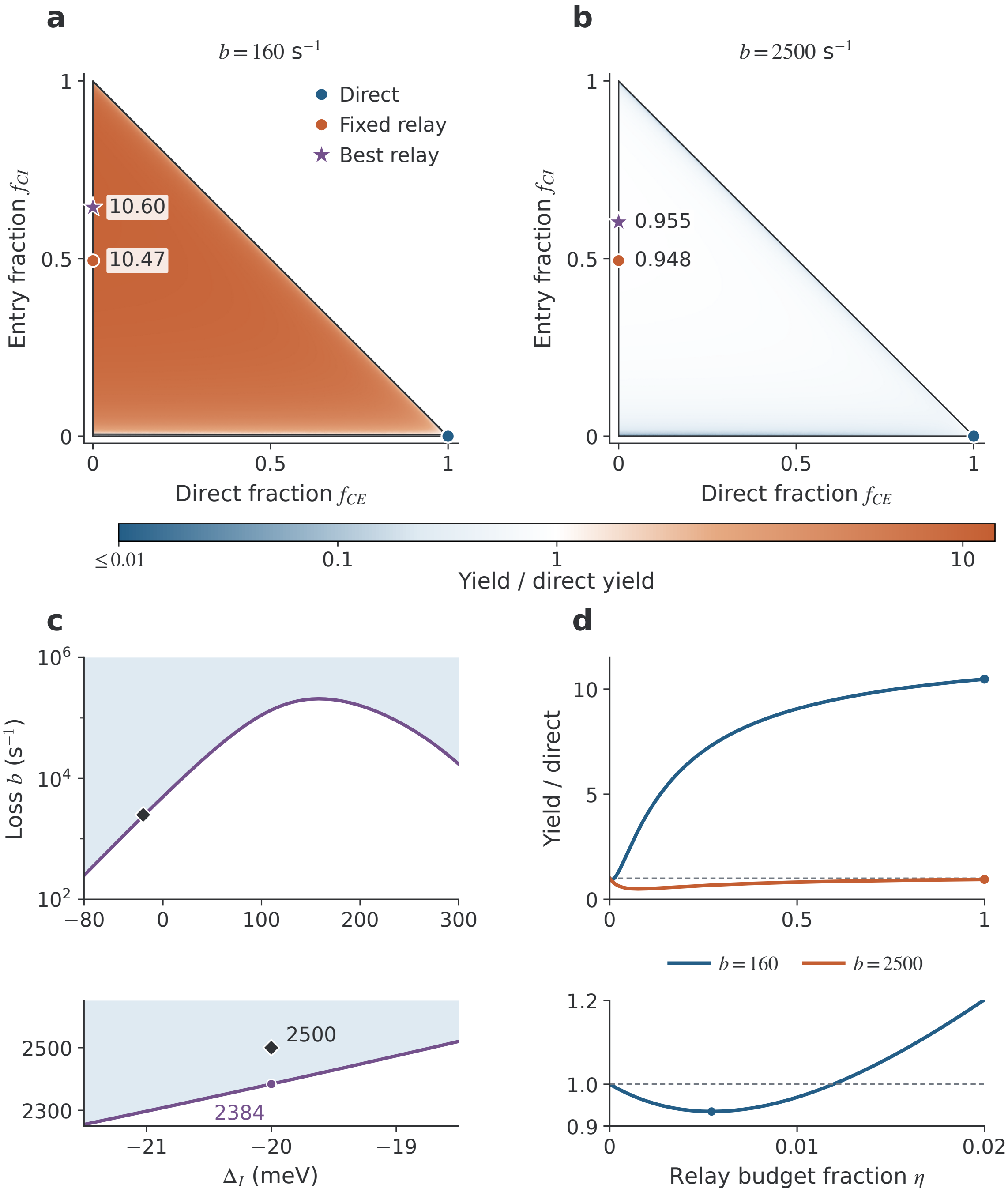

Figure 3: Above a critical intermediate loss, direct transfer maximizes total yield for a fixed squared-coupling budget and allocation-independent terminal rates. (a,b) Maps show $C$-prepared yield relative to direct transfer at $\Delta_I = -20$ meV, with $f_{IE} = 1 - f_{CE} - f_{CI}$. The black contour in (a) marks equal yield. The best pure relay (star) gives 10.60 and 0.955 times the direct yield at $b = 160$ and $2500\,\mathrm{s}^{-1}$, respectively. (c) Shading marks $b \geq b_*(\Delta_I)$. At $-20$ meV, the receiver benchmark (diamond, $b = 2500\,\mathrm{s}^{-1}$) lies above $b_* \simeq 2384\,\mathrm{s}^{-1}$. (d) At the same energy, both curves run from direct transfer ($\eta = 0$) to the fixed pure relay (circles at $\eta = 1$, $s = 0.49417$). The lower plot enlarges the initial dip below the direct yield (dashed line) at $b = 160\,\mathrm{s}^{-1}$.

They give the total photon yield

$$\Phi_C = \frac{rc_{\text{eff}}}{\kappa_{\text{eff}}(a_{\text{eff}} + c_{\text{eff}}) + a_{\text{eff}}d_{\text{eff}}}. \tag{19}$$

The time dependence of the emission still follows from the full generator.

For direct transfer, $\Phi_{C0} = rc_0/D_0$, with $d_0 = \chi_C c_0$ and $D_0 = \kappa(a+c_0)+ad_0$. Differentiating Eq. (19) at the direct endpoint gives

$$\partial_\eta \ln \Phi_C\big|_{\eta=0} = -\frac{\kappa a + (\kappa + d_0)p + (a + c_0)\chi_I h}{D_0} < 0, \qquad b > 0. \tag{20}$$

Reaching $E$ from $C$ through $I$ requires both relay links. For $b > 0$, their contribution $uw/\mathcal{B}$ to the forward rate enters at order $\eta^2$. Diverting coupling from the direct link and opening loss channels through $I$ already reduce yield at order $\eta$. A relay can therefore gain an advantage only after a finite fraction of the budget has shifted.

Figure 3(d) shows this dip at $b = 160\,\text{s}^{-1}$. For the fixed split, the yield falls about 6.5% below the direct value before rising above it. At $b = 0$, $\mathcal{B} = \eta(\nu + h)$, so the relay contribution itself becomes linear in $\eta$ when both relay links are present.

## Why the allocation endpoints suffice

The initial slope establishes a local cost. To compare yields along the entire allocation ray, write $\mathcal{A} = c\mathcal{B} + uw$ and define

$$\begin{aligned}\mathcal{K} &= \kappa u + \chi_C cu + \chi_I(a + c + u)w,\\ \mathcal{D} &= \kappa a\mathcal{B} + (\kappa + \chi_C a)\mathcal{A} + b\mathcal{K}.\end{aligned} \tag{21}$$

The yield is then $\Phi_C = r\mathcal{A}/\mathcal{D}$. After cross multiplication, its difference from the direct yield $\Phi_{C0}$ has the sign of

$$\mathcal{N} = \kappa a\Big[(c - c_0)\mathcal{B} + uw\Big] - c_0 b\mathcal{K}. \tag{22}$$

At fixed $s$, $\mathcal{N}$ is at most quadratic in $\eta$ and vanishes at $\eta = 0$. Factoring out $\eta$ gives an affine interpolation for $\eta > 0$,

$$\frac{\mathcal{N}(\eta)}{\eta} = (1-\eta)\mathcal{N}_0' + \eta\mathcal{N}(1), \tag{23}$$

where $\mathcal{N}_0' = \partial_\eta \mathcal{N}|_{\eta=0}$. For $b > 0$, Eq. (20) fixes the sign at the direct endpoint,

$$\mathcal{N}_0' = c_0 b D_0\, \partial_\eta \ln \Phi_C|_{\eta=0} < 0.$$

The interpolation can therefore become positive only if $\mathcal{N}(1) > 0$. At $b = 0$, the linear term in $\mathcal{N}$ vanishes and $\mathcal{N}(\eta) = \eta^2\mathcal{N}(1)$ gives the same conclusion. On each ray, an improving allocation exists if and only if the pure-relay endpoint beats direct transfer.

It remains to bound a mixture by the pure relay when that endpoint has the higher yield. Fix $s$ with $p, h > 0$. For this comparison, vary $c_0$ as an algebraic parameter while holding $p, h, \nu$, the thermal weights, and the terminal rates fixed. The pure-relay yield stays constant and the direct yield increases strictly with $c_0$. They coincide at

$$c_* = \frac{\kappa aph}{\kappa a(\nu + h) + b(a+p)(\kappa + \chi_I h)}. \tag{24}$$

For $c_0 \geq c_*$, the direct source matches or exceeds the pure relay, and Eq. (23) bounds every mixture by the direct yield.

For $0 \leq c_0 \leq c_*$, fix a mixture $0 < \eta < 1$. The pure relay matches or exceeds the yield of this mixture if $\mathcal{A}_1\mathcal{D}_\eta - \mathcal{A}_\eta\mathcal{D}_1 \geq 0$, where subscripts denote the relay fraction. Both $\mathcal{A}_1 = ph$ and $\mathcal{D}_1$ are independent of $c_0$, while $\mathcal{A}_\eta$ and $\mathcal{D}_\eta$ are at most linear in $c_0$. Thus the cross-multiplied difference is affine in $c_0$. We need only check its sign at $c_*$ and at zero.

At $c_0 = c_*$ the two endpoint yields coincide, and the preceding bound places every mixture at or below their common yield. At $c_0 = 0$ the direct link vanishes. All remaining internal rates scale with $\eta$, giving

$$\frac{rph}{\Phi_C(\eta, s)} = \frac{\kappa ab}{\eta^2} + \frac{\kappa a(\nu + h) + b(\kappa p + \chi_I ah)}{\eta} + (\kappa + \chi_C a + b\chi_I)ph. \tag{25}$$

All coefficients are nonnegative, and the $1/\eta$ coefficient is positive. The right-hand side decreases as $\eta$ grows, so the yield is largest at $\eta = 1$. The pure relay therefore bounds the mixture at both ends of the interval in $c_0$. Linear interpolation extends the bound to every $c_0$ between them, completing the proof of Eq. (18).

When a relay link vanishes, the same bound follows by restricting the dynamics to the accessible transient states (SI3.1). Mixtures can tie the best endpoint in degenerate cases, including $b = 0$ with $c_0 = c_*$.

## A loss threshold for every allocation

To find where the relay advantage disappears, set $\eta = 1$ in Eq. (22) and solve for the loss $b$ at equal yields. This gives the threshold $b_c(s)$ for a chosen split and the global threshold $b_*$ after optimization,

$$\begin{aligned} b_c(s) &= \frac{\kappa a\,[ph/c_0 - \nu - h]}{(a+p)(\kappa + \chi_I h)}, \\ b_* &= \max\left\{0, \max_{0 \le s \le 1} b_c(s)\right\}. \end{aligned} \tag{26}$$

For $0 \le b < b_*$, at least one pure relay outperforms direct transfer. For $b \ge b_*$, direct transfer maximizes total yield over the entire allocation simplex. The numerator and denominator of $b_c(s)$ are at most quadratic in $s$. The cubic terms cancel on differentiation, so the stationary points satisfy an equation of degree at most two. The admissible real roots and the two endpoints thus suffice to evaluate the maximum in Eq. (26).

At zero intermediate loss, the relay still has to offset the direct coupling it displaces. A given split succeeds when its effective transfer rate $ph/(\nu + h)$ exceeds $c_0$. The best such rate

is

$$\max_{0\le s\le 1} \frac{u_B w_B s(1-s)}{v_B s + w_B(1-s)} = \frac{u_B w_B}{(\sqrt{v_B}+\sqrt{w_B})^2}, \tag{27}$$

attained at $s = \sqrt{w_B}/(\sqrt{v_B}+\sqrt{w_B})$. Thus a relay can beat direct transfer at $b = 0$ exactly when

$$\frac{u_B w_B}{c_0} > (\sqrt{v_B}+\sqrt{w_B})^2.$$

If this condition fails, increasing $b$ preserves the direct optimum. The yield at each allocation is nonincreasing in $b$, and the direct yield stays fixed.

At $\Delta_I = -20$ meV and $b = 160\,\mathrm{s}^{-1}$, the best pure relay yields 10.60 times as many photons per preparation as direct transfer [Figure 3(a)]. As $b$ increases, the fixed split used in Figure 2 loses its advantage at $b_c(0.49417) = 2365.87\,\mathrm{s}^{-1}$. A different split can still outperform direct transfer between this crossing and the global threshold $b_* = 2383.86\,\mathrm{s}^{-1}$ [Figure 3(c)].

The receiver working point, $b = 2500\,\mathrm{s}^{-1}$, lies above both thresholds. Optimizing the split raises the relay-to-direct yield ratio from 0.9479 to 0.9548 [Figure 3(b)]. The best pure relay still falls short, and the endpoint bound establishes that direct transfer is globally optimal. Section 5 uses this reference to ask whether earlier emission from the fixed relay can reduce the preparation needed to activate the receiver and let it close on time.

# 5 Earlier emission lowers the preparation cost

At the working point of Section 4 ($\Delta_I = -20$ meV, $b = 2500\,\mathrm{s}^{-1}$), the fixed relay emits 5.21% fewer photons per preparation than direct transfer. The mean emission time falls from 7.69 to 1.07 ms [Figure 4(a)]. We first compare their receiver responses at equal preparation. Then we find the preparation each source needs to meet the same activation and closure targets.

## From photon arrival to receiver response

The photon yield collected by time $W$ is

$$F_W = \int_0^W I_C(t)\,dt = r\boldsymbol{e}_E^{\mathsf{T}}(-L)^{-1}(\mathbf{I}_3 - e^{LW})\boldsymbol{e}_C. \tag{28}$$

Here $\mathbf{I}_3$ is the identity on the $(C, I, E)$ subspace. Direct transfer gives the highest yield over sufficiently short windows. Its leading term is $rcW^2/2$, while the two transfers in a pure relay give $ruwW^3/6$. By $W = 1$ ms the fixed relay has delivered 13.96 times as many photons as the direct source. More generally, a mixed allocation can outperform both pure routes before a finite deadline (SI3). The receiver response depends on which wavelengths activate the channel and how long it stays open.

Both sources emit from the same thermally relaxed state $E$. Their common photon spectrum $\rho_\gamma(\lambda)$ is time independent, integrates to one over wavelength, and has a calculated peak at 594 nm and a full width at half maximum of 135 nm. We use the four-state photocycle of Gupta et al.[77] with their shared parameters and the Chrimson closing rate $G_{d1} = 41\,\mathrm{s}^{-1}$. The relative conductance is $g = O_1 + 0.05O_2$, where $O_1$ and $O_2$ are the populations of the two conducting states.

We measure effective preparation by $q = B_C\Phi_{C0}$, where $B_C$ combines the prepared population with a common, spectrally flat optical coupling.

The action spectrum $\mathcal{A}_j$ weights the photons that drive transition $j$, giving

$$\xi_j = \int_0^\infty \rho_\gamma(\lambda)\frac{\mathcal{A}_j(\lambda)}{\mathcal{A}_j(594\,\mathrm{nm})}\,d\lambda, \qquad x_j(t) = \frac{\xi_j q I_C(t)}{k_1\Phi_{C0}}, \tag{29}$$

where $\xi_j$ is the spectral factor, $x_j(t)$ the transition drive, and $k_1 = 3000\,\mathrm{s}^{-1}$. We sum photon contributions across wavelengths before applying the saturation law to the broadband input. For the first activation channel, we use the measured initial-photocurrent action spectrum of Chrimson–EYFP as a proxy.[78] It gives $\xi = 0.61134$. The baseline model assigns this factor to all light-dependent transitions. Channel-specific spectral tests are detailed in SI2.2.

A reduced three-state receiver uses the same initial excitation scale at weak flux. Its opening drive $\alpha(t)$ and integrated dose $D$ are

$$\alpha(t) = \frac{\xi q I_C(t)}{\Phi_{C0}}, \qquad D = \int_0^\infty \alpha(t)\,dt = \xi q \frac{\Phi_C}{\Phi_{C0}}. \tag{30}$$

Equal $q$ thus means equal effective preparation before spectral weighting. The relay's smaller yield gives it a smaller dose. The available, open, and refractory populations $(A, O, R)$ obey

$$\dot{O} = \alpha A - \gamma O, \qquad \dot{R} = \beta\gamma O - k_r R, \qquad A + O + R = 1, \tag{31}$$

with $A(0) = 1$. We set $\gamma^{-1} = 15.8$ ms using ChrimsonR as a closing-time reference[79] and choose $\beta = 0.2$ and $k_r = 10\,\mathrm{s}^{-1}$ separately. The full photocycle equations, parameter sources, and optical units are given in SI2.

Write $J_W^y = \int_0^W y(t)\,dt$ for the integrated response, with $y = g$ or $O$. At $q = 1$, the four-state mean conductance over 20 ms is 0.2074 for direct transfer and 0.2358 for the relay, a 13.68% gain [Figure 4(b)]. Figure 4(c) locates this gain in the energy–loss plane. At $\Delta_I = -20$ meV, direct transfer maximizes total yield over all allocations for $b \geq b_* = 2383.86\,\mathrm{s}^{-1}$. The fixed relay still gives the larger 20 ms conductance integral until $b \simeq 2985\,\mathrm{s}^{-1}$. Throughout this interval, changing the coupling allocation cannot improve total yield, yet the fixed relay produces the larger early response.

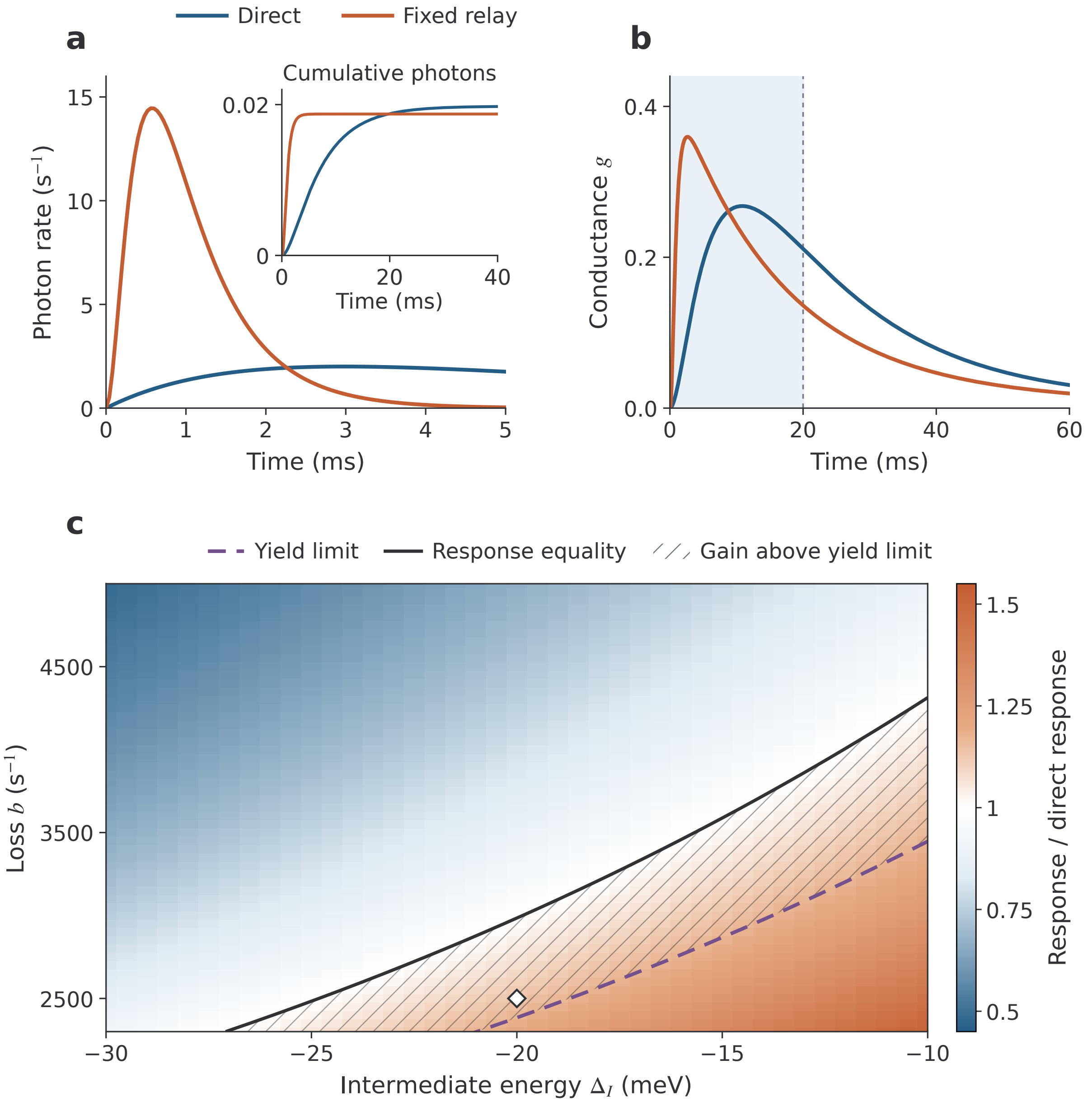

Figure 4: At equal preparation, a lower-yield relay can drive a larger early receiver response. (a) The fixed relay (orange, $s = 0.49417$) emits earlier than the direct source (blue) but produces 5.21% fewer photons per preparation (inset, cumulative yield). (b) At effective preparation $q = 1$, the relay increases integrated conductance of the four-state Chrimson receiver by 13.68% over the shaded 20 ms window. Both sources share a calculated spectrum peaking at 594 nm and a spectral factor $\xi = 0.61134$. (c) Colors give the relay/direct ratio of the 20 ms conductance integrals. In the hatched band above the dashed yield threshold and below the solid response-equality curve, direct transfer maximizes yield and the fixed relay drives the larger response. The diamond marks the conditions in (a,b), $\Delta_I = -20$ meV and $b = 2500\,\mathrm{s}^{-1}$.

## The preparation required for activation and closure

We now hold the receiver task fixed. A single value of $q$ must produce a mean response of at least $y_*$ over $0 \le t \le W$ and leave a residual response of at most $\epsilon_c$ at $T_c > W$. For a specified source, the minimum effective preparation is

$$q_{\min} = \inf \left\{ q \ge 0 : \frac{J_W^y(q)}{W} \ge y_*, \quad y(T_c; q) \le \epsilon_c \right\}, \tag{32}$$

with the receiver initially dark. The closure criterion measures residual conductance or open population at the deadline. Other receiver populations may still be recovering.

In Figure 5, the target mean is 0.20 over the first 20 ms, a window on the scale of channel closing. The residual must fall to 0.05 or less by 50 ms. In the four-state model, the numerical minimum falls from $q = 0.95556$ for direct transfer to 0.79753 for the relay, a 16.54% saving. At matched mean activation, the respective residual conductances are 0.04653 and 0.02482 [Figure 5(b)]. Activation sets both minima. The relay also leaves a larger margin for closure. The three-state receiver gives a preparation saving of 20.14% [Figure 5(c)].

Tightening the residual limit to 0.03 changes which sources can meet the task. The relay still satisfies both targets. For direct transfer, the numerical search finds no effective preparation satisfying both targets over $10^{-5} \le q \le 10^3$ [Figure 5(a)]. Search details, additional thresholds, and independent trajectory checks are given in SI4.

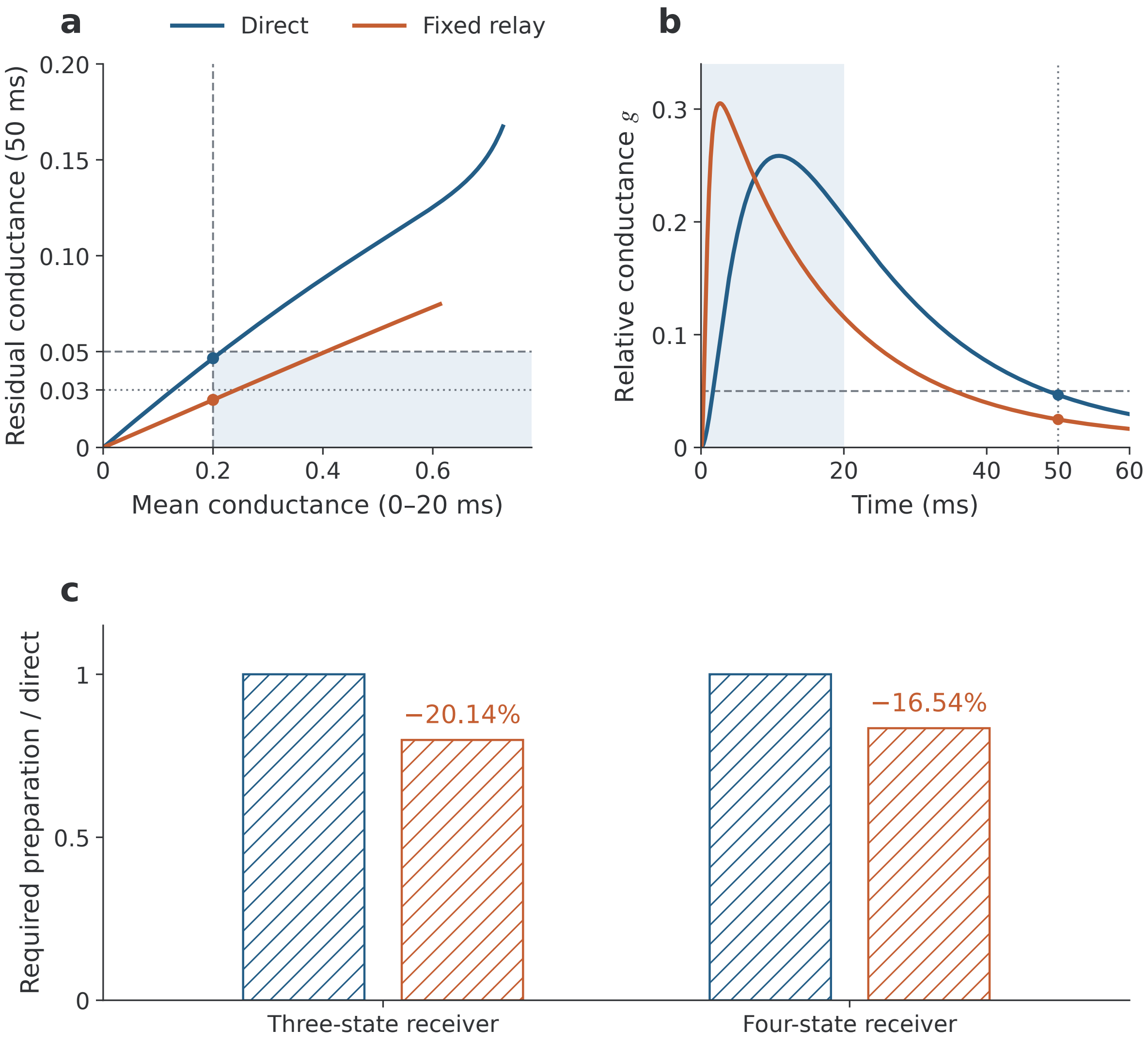

Figure 5: The fixed relay reaches the same mean activation with less preparation and leaves less residual conductance. (a) For the four-state receiver, shading marks mean conductance $\geq 0.20$ over 0–20 ms and residual conductance $\leq 0.05$ at 50 ms. Circles mark the minimum effective preparation $q$ for each source. Both curves end at $q = 30$, where the direct source gives the larger mean. With the tighter residual limit of 0.03 (dotted), only the relay meets both targets in the search $10^{-5} \leq q \leq 10^{3}$. (b) At matched mean conductance, residuals at 50 ms are 0.0465 for direct transfer (blue) and 0.0248 for the relay (orange). (c) Relative to direct transfer, the relay reduces the required preparation by 20.14% in the three-state model and 16.54% in the four-state model.

The brief four-state response is dominated by the first activation channel. With a common spectral factor, the drive depends on $\xi q$. Spectral weighting therefore multiplies each monochromatic-reference cost by $1/\xi$ and preserves the percentage saving. To test the common-factor assumption, we hold the first-channel factor fixed and independently set each remaining factor to 0.5, 1, or 2 times that value. Across these 27 combinations, the saving ranges from 16.532% to 16.544% for the brief task starting from darkness (SI2.2).

## How photon timing governs activation and closure

The reduced receiver lets us assign a contribution to each photon arrival time. Let $h_C(t) = I_C(t)/\Phi_C$ denote the emission-time density conditional on a photon being emitted. At weak drive, an integrating factor in Eq. (31) gives

$$O(t) = D \int_0^t h_C(u) e^{-\gamma(t-u)} \, du + \mathcal{O}(D^2). \tag{33}$$

An opening initiated at time $u$ contributes to the activation integral until $W$ and to the residual if it persists until $T_c$. Integrating the response over the window and evaluating it at the deadline give the two weights

$$\begin{aligned} \mathcal{C}_W &= \int_0^W h_C(u)[1 - e^{-\gamma(W-u)}] \, du, \qquad \gamma J_W^O = D\mathcal{C}_W + \mathcal{O}(D^2), \\ \mathcal{H}_{T_c} &= \int_0^{T_c} h_C(u) e^{-\gamma(T_c-u)} \, du, \qquad\qquad O(T_c) = D\mathcal{H}_{T_c} + \mathcal{O}(D^2). \end{aligned} \tag{34}$$

A photon arriving early has time to contribute to activation and still allow the channel to close. A photon arriving just before $T_c$ carries a residual weight close to one and arrives too late to contribute to activation.

The capture weight also gives an exact bound for the reduced receiver. Since $A \leq 1$, replacing $A$ by unity in the opening term $\alpha A$ bounds $O(t)$ from above at any drive strength. Thus

$$q \geq \frac{\Phi_{C0} \gamma W y_*}{\xi \Phi_C \mathcal{C}_W} \tag{35}$$

is necessary to reach the activation target. The product $\xi\Phi_C\mathcal{C}_W$ combines spectral sensitivity, photon yield, and the temporal weight for capture within the window.

At weak drive, activation sets a lower dose and closure sets an upper dose. For compatibility to leading order, we require

$$\frac{\gamma W y_*}{\mathcal{C}_W} \leq D \leq \frac{\epsilon_c}{\mathcal{H}_{T_c}}, \qquad \frac{\gamma W y_* \mathcal{H}_{T_c}}{\mathcal{C}_W} \leq \epsilon_c, \tag{36}$$

when both weights are positive. The yield converts dose to preparation through $D = \xi q\Phi_C/\Phi_{C0}$. It cancels from the second inequality, which tests whether the emission times are compatible with both targets. Figure 5 makes the corresponding comparison using the full receiver dynamics.

The preparation ranking can be tested by replaying the complete emission waveforms in the same voltage-clamped Chrimson-expressing cells. Each trial would start from the same dark-adapted state and use a common spectrum and optical calibration. Varying the input scale while preserving each waveform's shape and photon yield per preparation would identify the least input that meets both conductance targets. Calibrated initial excitation numbers relate this input to preparation cost. The measured waveforms and spectra can also drive a characterized receiver model. Testing the source yield–time relation or the allocation threshold requires the state and rate calibrations in SI1.4. An equal-incident-fluence comparison is given in SI6.

# 6 Protocol dependence of the timing advantage

Does the fixed relay retain its advantage when the protocol changes? Figure 6 compares the same sources and four-state receiver as we vary the response window, effective preparation, preparation duration, and pulse count.

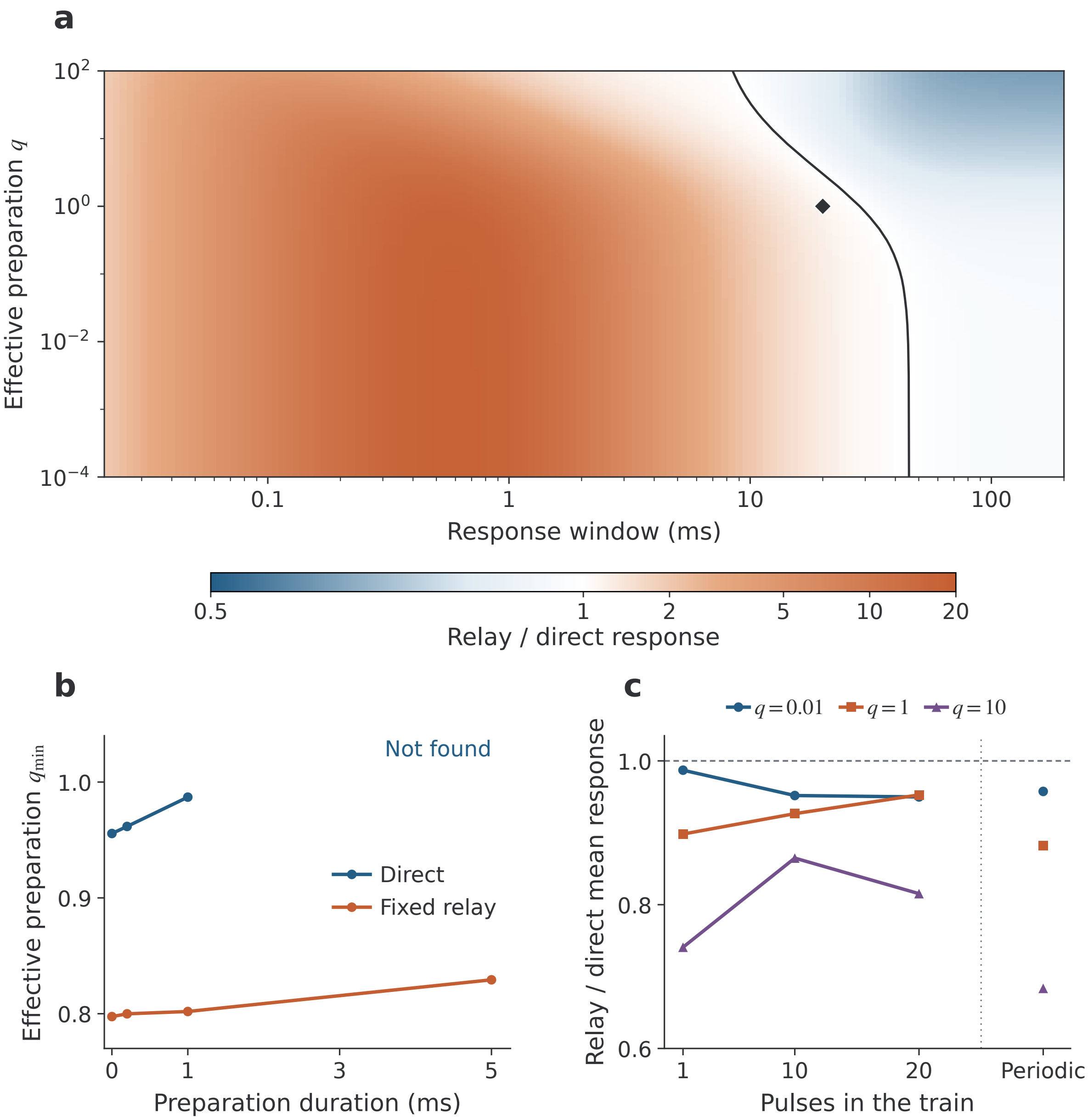

Figure 6: Longer response windows or sustained stimulation can favor direct transfer over the early-emitting relay. (a) The black contour marks equal integrated conductance in the four-state Chrimson model. At equal preparation $q = 1$, the 20 ms window favors the relay (diamond). Direct transfer gives the larger integral beyond about 28.4 ms. (b) Minimum $q$ for the activation and closure targets of Figure 5 as preparation is spread uniformly over the indicated duration. Zero denotes an impulse. At 5 ms, only the relay meets both targets in the search $10^{-5} \leq q \leq 10^{3}$. (c) Trains of $N$ pulses spaced by 50 ms start from a dark receiver. Direct transfer gives the larger mean conductance over the full $50N$ ms at each effective preparation shown, as it does at periodic steady state (separate symbols).

## Response window, effective preparation, and pulse duration

At equal preparation $q = 1$, the relay gives the larger conductance integral over 20 ms. Extending the window past about 28.42 ms reverses the ranking [Figure 6(a)]. At fixed $W = 20$ ms, the ranking also reverses as the effective preparation increases past $q \simeq 3.016$. Saturation and redistribution among receiver states limit the benefit of early photons. At $q = 30$ the direct source has the larger mean, explaining the unequal horizontal endpoints in Figure 5(a). Controls for the reduced receiver and the closing rate appear in SI6.

A finite preparation pulse changes the photon waveform itself. For uniform preparation over a duration $t_p > 0$, the emitted rate is $\widetilde{I}_C = s_p * I_C$, with $s_p(t) = 1/t_p$ on $0 \le t \le t_p$ and zero elsewhere. This spreads emission over time while preserving the total preparation and photon yield. Figure 6(b) applies the activation and closure targets of Figure 5 to these broadened pulses. At $t_p = 1$ ms, the relay saves 18.75% in preparation. At $t_p = 5$ ms, only the relay meets both targets within the search range $10^{-5} \le q \le 10^3$.

## Repeated stimulation and sustained opening

For repeated stimulation, we ask which source sustains the larger mean conductance at the same preparation per pulse. For $N$ preparations spaced by $T$, Figure 6(c) uses the mean over the full train,

$$\overline{y}_{\text{train},N} = \frac{1}{NT} \int_0^{NT} y(t)\, dt, \tag{37}$$

starting with a dark receiver and no prior source excitation. For a single pulse at $q = 1$, averaging over the full $T = 50$ ms already favors direct transfer. At the same preparation per pulse, the relay/direct ratio is 0.927 over ten periods and 0.882 in the periodic steady state. Last-period averages and a control that removes source startup further separate the effects of averaging, source history, and receiver adaptation (SI6). The train calculations use the shared spectral factor. Extending them to light-adapted Chrimson requires a spectral response characterized for the relevant illumination history.[80]

For the reduced receiver, sustained opening admits an exact upper bound. Consider a bounded, nonnegative periodic drive with dose $D = \int_0^T \alpha(t)\,dt$ in each period. For $\gamma, k_r > 0$ and $0 \leq \beta \leq 1$, the mean open population at periodic steady state satisfies

$$\langle O \rangle \leq \frac{D}{\gamma T + D(1 + \beta\gamma/k_r)}. \tag{38}$$

Constant illumination attains the bound. The proof retains the correlation between drive $\alpha$ and availability $A$ throughout the cycle.

For $\beta > 0$, periodic balance gives $\langle R \rangle = (\beta\gamma/k_r)\langle O \rangle$. Define $c_r = 1 + k_r/(\beta\gamma)$, $d_r = 1 - c_r R$, $\ell = (1-\beta)\gamma$, and $F_r = \ell(1-R) + k_r R$. The periodic solution for the refractory population obeys

$$R(t) = \int_0^\infty \beta\gamma[1 - A(t-u)]e^{-(\beta\gamma + k_r)u}\,du < \frac{\beta\gamma}{\beta\gamma + k_r}, \tag{39}$$

so $d_r > 0$. The population equations now give $A = d_r - \dot{R}/(\beta\gamma)$ and $\dot{A} = F_r - (\alpha + \ell)A$. Over a period, $\dot{A}/A$ averages to zero, giving $\langle \alpha \rangle = \langle F_r/A \rangle - \ell$.

Convexity of the reciprocal gives $1/A \geq 1/d_r + \dot{R}/(\beta\gamma d_r^2)$. After multiplication by $F_r$, the term containing $\dot{R}$ averages to zero because it is a total derivative of a function of $R$. Applying Jensen's inequality to $R/(1 - c_r R)$ then gives

$$\langle \alpha \rangle \geq \frac{k_r}{\beta}\left\langle \frac{R}{1 - c_r R} \right\rangle \geq \frac{\gamma\langle O \rangle}{1 - (1 + \beta\gamma/k_r)\langle O \rangle}. \tag{40}$$

Substituting $\langle \alpha \rangle = D/T$ yields Eq. (38). For $\beta = 0$, the periodic state has $R = 0$. Jensen's inequality applied to $\langle \alpha \rangle = \gamma(\langle 1/A \rangle - 1)$ gives the same bound.

At equal preparation per period, the lower-yield relay delivers a smaller dose $D = \xi q \Phi_C/\Phi_{C0}$ and therefore has a lower ceiling on sustained opening in the reduced model. In the brief task, the same early photons can contribute to activation before $W$ and still leave time for closure before $T_c$. For sustained stimulation, the four-state calculations instead favor direct transfer across the pulse trains shown in Figure 6(c).

# 7 Conclusions and Outlook

Above a sharp threshold in intermediate loss, direct transfer maximizes total photon yield. This result holds across all allocations of a fixed squared-coupling budget when terminal losses are independent of allocation. At the high-loss working point of the spectrally weighted Chrimson benchmark, the fixed relay emits 5.21% fewer photons per prepared excitation than direct transfer but delivers them earlier. It reaches the same mean activation with 16.54% less effective preparation and leaves lower residual conductance at the deadline. When sustained opening is the goal, repeated stimulation can favor direct transfer again.

For an experimentalist observing a brighter mechanoluminescent pulse, the mean photon count $N_C q_C \Phi_E$ identifies three possible origins of the gain. The material may prepare more excitations, deliver a larger fraction to the emitter, or emit more efficiently after arrival. Paired mechanical and optical preparations separate these effects once the initial populations are calibrated. The yield ratio $\Phi_C/\Phi_E$ gives the probability of reaching the emitter. The difference $\bar{t}_C - \bar{t}_E$ gives the mean time taken by the excitations that reach it. For a reversible source with one intermediate and positive intermediate loss, the two yields also predict the PL mean emission time if the terminal rates and thermal weights are known independently. A separate optical decay trace can test that prediction.

A single ZnS:$Mn^{2+}$ particle offers a concrete setting for this test. Mukhina and co-workers observed mechanoluminescence in regions with weak Mn photoluminescence. They proposed two routes to Mn excitation, one through local carrier recombination and another through a Mn-bound exciton.[72] A paired measurement would record the Mn emission band from one region after mechanical preparation and again after selective optical excitation of Mn. Once excitation populations, pulse profiles, and collection efficiencies are calibrated, a temperature series can track changes in entry probability and delay. Electronic-structure calculations can estimate how donor–Mn separation and local strain affect the energy gaps, lattice relaxation, and couplings along each route. At each temperature, one set of rates should account for both emission traces. A transient signal assigned independently to the bound exciton would test whether it accumulates during the rise of Mn emission.

A weak optical probe could map how activation and residual conductance depend on photon arrival time. A short, weak probe of known fluence would be applied at different delays in separate trials on dark-adapted, voltage-clamped Chrimson-expressing cells. The added conductance integrated over the activation window measures the contribution to opening. The added conductance at the deadline measures the contribution to residual activity. The two fluence-normalized signals give empirical response kernels as functions of probe delay. After calibration to the reduced receiver, their averages over the normalized emission waveform give the source-specific weights in Eq. (34). Over what range of drive strengths can these weak-probe measurements predict the source ranking? Repeating the probe during a pulse train, with an identical unprobed train as the reference, would test how the responses change with illumination history.[80]

The next theoretical step is to find the transfer allocation that minimizes preparation for the receiver task itself. For total yield, a pure pathway always attains the optimum. For collection within a finite window, a mixed pathway can outperform both pure routes. At weak drive, the capture and closure weights map each allowed source waveform to a mean activation and a deadline residual. The boundary of this set would give the greatest activation attainable at each residual limit and preparation level. Two further measurements would bring that prediction into material design: the excitations generated per unit mechanical work and the fraction of emitted light reaching the opsin at each wavelength.

When yields and mean times coincide, which measurement distinguishes dark states in series from dark states in parallel? When does a mixed pathway meet both receiver targets at lower cost than either pure route? Can single-pulse kinetics predict when a pulse train will reverse the source ranking? Ultimately, can we predict before synthesis which defect modification will lower the mechanical work per successful optogenetic stimulus? These questions guide our ongoing work.

# Acknowledgement

J.C. gratefully acknowledges Dr. Puxian Xiong and Dr. Changhong Linghu for insightful discussions and valuable suggestions.

# Supporting Information Available

Source and receiver parameters; state reduction, thermal weights, and a qualitative ZnS:$Mn^{2+}$ correspondence; preparation normalization and calibration; emission-spectrum construction, experimental action-spectrum projection, and channel-factor tests; allocation boundary limits, short-time photon collection, and a finite-window counterexample; activation–closure task searches; numerical methods; photon-window, equal-fluence, reduced-receiver, and history controls.

# Competing Interests

The authors declare no competing financial interests.

# Data Availability Statement

The experimental Chrimson action spectrum is from the published source data for Figure 1D of Govorunova et al.[78] The simulated emission waveforms, receiver responses, and numerical data underlying the figures are available from the corresponding authors upon reasonable request.

# Supporting Information: Nonemissive Intermediates: Photon-Yield Limits and Timing Advantages for Optogenetic Light Sources

Jingqi Chen,[*,†] Qiang Zhou,[*,‡,†] Yue Zhu,[†] and Xin Xu[*,¶,§]

† *Starlight Origin Technology Co., Ltd., Shenzhen, China*

‡ *Department of Mechanical Engineering, City University of Hong Kong, Hong Kong SAR, China*

¶ *Department of Materials Science and Engineering, National University of Singapore, Singapore*

§ *School of Materials Science and Engineering, Harbin Institute of Technology, Harbin, China*

E-mail: bestcjq@gmail.com; qzhou229-c@my.cityu.edu.hk; xin.xu@nus.edu.sg

## SI1 Source parameters and comparison families

The source generator, observables, and productive budget are defined in Eqs. (1)–(4); the microscopic rate realization is specified below. Table S1 records the original kinetic realization; Table S3 gives the optically shifted realization used for the receiver. The coordinate $x$ is dimensionless. Energies use $\epsilon_G = 0$, and all rates used in propagation are in $s^{-1}$. Table S2 specifies the comparison families and parameter scans.

All source observables use unit initial excitation before receiver exposure is applied. The direct reference is independent of $b$ because $I$ is decoupled.

Table S1: Original microscopic realization of the source kinetics. The receiver uses the kinetically equivalent optical realization in Table S3; all unlisted parameters remain the same.

| Quantity | Value |
|---|---|
| $T_b$, $K_x$ | 300 K, 2.4 eV |
| $(x_G, x_C, x_I, x_E)$ | $(1.8, 0, 0.62, 1)$ |
| $f_q$, $\hbar\omega$ | 0.10, 0.090 eV |
| $\epsilon_E$ | 2.511800259 eV |
| $\epsilon_C - \epsilon_E$ | 0.3061997412 eV |
| $\Delta_I$ at the central point | $-0.020$ eV |
| $V_B^2$ | $0.5144\,(\mu\text{eV})^2$ |
| $V_{CG}$; resulting $a$ | $1.0\,\mu$eV; $574.982174706\,\text{s}^{-1}$ |
| $r$, $n$ | 125, 41.66666667 $\text{s}^{-1}$ |
| $\chi_C$, $\chi_I$ at $-20$ meV | $7.1791122934 \times 10^{-6}$, 2.1676302962 |
| $c_0$, $u_B$, $w_B$ at $-20$ meV | 15.5746057094, 8794.6578564454, 2723.1889895207 $\text{s}^{-1}$ |

Table S2: Source families used in the main and supporting figures.

| Family | Definition and scope |
|---|---|
| Direct reference | $\eta = 0$, full productive budget on $CE$; $I$ is decoupled. |
| Frozen task relay | $\eta = 1$, $s = 0.4941709511604826$, $\Delta_I = -20$ meV, $b = 2500\,\text{s}^{-1}$. Used in Figures 1 and 4–6, and in the receiver controls in SI6. |
| Classification | Same pure-relay split; $\Delta_I \in [-80, 300]$ meV and independently controlled loss $b \in [10^{-3}, 10^{6}]\,\text{s}^{-1}$ in Figure 2a; $V_{IG}$ is adjusted when energy varies at fixed $b$. Figure 2(d) separately uses $\Delta_C = 20$ meV, $\Delta_I = -20$ meV, and $b = 150\,\text{s}^{-1}$. |
| Allocation family | Full $0 \le \eta, s \le 1$ at fixed energy and terminal losses in Figure 3. |

## SI1.1 Microscopic rate realization

For the numerical realization, we use one nondegenerate surface per state with common internal partition factors, so that $\chi_i = \exp[-(\epsilon_i - \epsilon_E)/(k_B T_b)]$. We parameterize the transfer rates using displaced harmonic surfaces,

$$U_i(x) = \epsilon_i + \frac{K_x}{2}(x - x_i)^2, \qquad \lambda_{ij} = \frac{K_x}{2}(x_i - x_j)^2, \tag{S1}$$

where $x$ is a collective nuclear coordinate, $\epsilon_i$ and $x_i$ are the energy and position of each minimum, $K_x$ is the common curvature, and $\lambda_{ij}$ is the reorganization energy. We use a thermally populated Marcus–Jortner rate,[S1,S2]

$$k_{i\to j} = \frac{2\pi |V_{ij}|^2}{\hbar\sqrt{4\pi\lambda_s k_B T_b}} \sum_{\ell\in\mathbb{Z}} W_\ell \exp\left[-\frac{(\epsilon_j - \epsilon_i + \lambda_s + \ell\hbar\omega)^2}{4\lambda_s k_B T_b}\right]. \tag{S2}$$

Here $V_{ij}$ is the electronic coupling. A fraction $f_q$ of the reorganization energy is assigned to a quantized mode of energy $\hbar\omega$, leaving $\lambda_s = (1 - f_q)\lambda_{ij}$ in the classical component. With $S = f_q\lambda_{ij}/(\hbar\omega)$ and $n_B = [\exp(\hbar\omega/k_BT_b) - 1]^{-1}$, the sideband weights are[S3]

$$W_\ell = e^{-(\mu_+ + \mu_-)} \left(\frac{\mu_+}{\mu_-}\right)^{\ell/2} \mathrm{I}_{|\ell|}(2\sqrt{\mu_+\mu_-}), \quad \mu_+ = S(n_B + 1), \quad \mu_- = Sn_B, \tag{S3}$$

where $\mathrm{I}_{|\ell|}$ is the modified Bessel function of the first kind of order $|\ell|$. The weights are nonnegative, sum to one, and obey $W_{-\ell} = e^{-\ell\hbar\omega/(k_BT_b)}W_\ell$. Pairing opposite sidebands gives the forward/reverse detailed-balance ratio.

## SI1.2 State reduction, thermal weights, and material applicability

**Closed populations and resolved outputs.** Let $\mathcal{B}_i$ collect the microstates assigned to $i = C, I, E$, with $p_i = \sum_{\alpha\in\mathcal{B}_i} p_{i\alpha}$. Starting from incoherent, time-homogeneous microscopic rate dynamics, exact closure for arbitrary initial populations is obtained if

$$\sum_{\beta\in\mathcal{B}_j} k_{i\alpha\to j\beta} = K_{i\to j} \quad \text{for every } \alpha \in \mathcal{B}_i, \quad j \neq i, \tag{S4}$$

with the same condition for each resolved terminal channel, including monitored emission. Summing the microscopic equations then gives the closed generator and $I(t) = rp_E(t)$.

Approximate closure follows from rapid intragroup thermalization when the mixing rate $\Gamma_i$ far exceeds all relevant escape rates and bath memory is short. After the mixing transient, $p_{i\alpha}(t) \simeq \rho_{\alpha|i}p_i(t)$.[S4] Both prepared populations and arrivals must reach this conditional distribution before appreciable transfer, emission, or loss. This requires population equilibration in addition to dephasing. For early-waveform predictions, the mixing transient must contribute negligibly on the observation scale or be included explicitly. The effective-generator identities are exact; their microscopic accuracy is controlled by this time-scale separation.

**Inheritance of internal detailed balance.** For microstate energies $E_{i\alpha}$ at a fixed bath temperature, write

$$Z_i = \sum_{\alpha\in\mathcal{B}_i} e^{-\beta_b E_{i\alpha}}, \qquad \rho_{\alpha|i} = \frac{e^{-\beta_b E_{i\alpha}}}{Z_i}, \qquad \beta_b = (k_BT_b)^{-1}. \tag{S5}$$

The sum denotes the restricted partition function, with integrals replacing sums for continuous degrees of freedom. Suppose the internal microstate rates obey $e^{-\beta_b E_{i\alpha}}k_{i\alpha\to j\beta} = e^{-\beta_b E_{j\beta}}k_{j\beta\to i\alpha}$. The conditional thermal average gives

$$K_{i\to j} = \sum_{\alpha\in\mathcal{B}_i} \rho_{\alpha|i} \sum_{\beta\in\mathcal{B}_j} k_{i\alpha\to j\beta}, \qquad Z_iK_{i\to j} = Z_jK_{j\to i}. \tag{S6}$$

The last equality follows by interchanging the two microstate sums. For exact lumpability the average equals the common exit rate; for fast thermalization it is the leading effective rate. With $\chi_E = 1$ and $\chi_i = Z_i/Z_E$, it gives precisely $d = \chi_C c$, $v = \chi_C u/\chi_I$, and $z = \chi_I w$. Internal detailed balance applies to transfer during post-preparation relaxation at fixed bath conditions; termination remains irreversible.

**Triplet multiplicity and microscopic rate laws.** If $Z_i = g_i Z_{\rm int} e^{-\beta_b \epsilon_i}$ with a common internal factor, then

$$F_i^{\rm th} - F_E^{\rm th} = \epsilon_i - \epsilon_E - k_B T_b \ln(g_i/g_E), \qquad \chi_i = \frac{g_i}{g_E} e^{-\beta_b(\epsilon_i - \epsilon_E)}. \tag{S7}$$

For three degenerate triplet sublevels and one singlet, $g_I/g_E = 3$. An exact grouping results when each triplet sublevel has the same exits to $C$, $E$, and each loss channel. For a per-sublevel return rate $w$ to $E$, microscopic detailed balance gives $E \to I_m$ the rate $we^{-\beta_b(\epsilon_I - \epsilon_E)}$; summation yields $z = 3we^{-\beta_b(\epsilon_I - \epsilon_E)}$. Rapid intragroup thermalization gives an approximate grouping when the exits differ. Split sublevels enter through their full partition sum.

At 300 K, $k_B T_b \ln 3 \simeq 28.4$ meV. The $\epsilon_i$ in Eq. (S2) remain potential minima: microscopic rates are calculated from these energies and their couplings before the channels are grouped. Multiplicity enters through the channel sum and thermal weights, rather than a substitution of free-energy gaps into an unchanged microscopic rate formula.

**Material-specific limits.** Defect or redox configurations require dilute single-excitation relaxation without continuing mechanical injection or broad unresolved trapping times. Triplets and radical-ion pairs require equivalent exits or rapid thermal equilibration of the grouped spin and spatial configurations; unresolved spin-selective recombination or coherent evolution requires additional dynamics.[S5,S6] Sensitizer migration can fit the network when it carries one excitation with closed population kinetics; multiphoton upconversion, excitation pooling, and dispersive migration do not satisfy that reduction.[S7]

For the allocation theorem, this reduction must hold throughout the budget domain, with productive rates linear in their allocated fractions and thermal weights, intragroup distributions, and terminal rates fixed. Material-specific calibration is described in SI1.4.

## SI1.3 An optical realization with a 594 nm photon-spectrum peak

We specify the $E \to G$ spectrum by a thermal Franck–Condon construction:[S3] equilibration on $E$ precedes emission, the transition dipole is independent of nuclear coordinates (Condon approximation), and the photon density of states is that of vacuum. Both sources emit solely from the same relaxed state in the same environment, giving the common time-independent spectrum and factorized photon rate $I_C(t)\rho_\gamma(\lambda)$.

With $\Delta\epsilon_{EG} = \epsilon_E - \epsilon_G$, $\lambda_{EG} = K_x(x_E - x_G)^2/2$, and the sideband weights of Eq. (S3) evaluated for this transition, the thermal Franck–Condon density is

$$\mathcal{D}_{EG}(E_\gamma) = \frac{1}{\sqrt{4\pi\lambda_s k_B T_b}} \sum_{\ell\in\mathbb{Z}} W_\ell \exp\left[-\frac{(E_\gamma - \Delta\epsilon_{EG} + \lambda_s + \ell\hbar\omega)^2}{4\lambda_s k_B T_b}\right], \tag{S8}$$

where $\lambda_s = (1 - f_q)\lambda_{EG}$. Spontaneous photon emission adds the $E_\gamma^3$ factor. For positive photon energies,

$$S_E(E_\gamma) = \frac{E_\gamma^3 \mathcal{D}_{EG}(E_\gamma)}{\int_0^\infty E^3 \mathcal{D}_{EG}(E)\, dE}, \qquad \rho_\gamma(\lambda) = S_E(hc/\lambda)\frac{hc}{\lambda^2}. \tag{S9}$$

Thus $\int \rho_\gamma \, d\lambda = 1$. The quoted spectral peak is defined by photon number per unit wavelength. The wavelength Jacobian makes its location equivalent to maximizing $E_\gamma^5 \mathcal{D}_{EG}(E_\gamma)$.

A common energy shift of $C, I, E$ relative to $G$ changes the optical gap while leaving all internal energy differences, reorganization energies, transfer rates, and detailed-balance weights unchanged. We retain the prescribed terminal rates by adjusting the $CG$ and $IG$ couplings for $a$ and $b$, and the effective radiative dipole for $r$, while keeping $n$ fixed. These quantities lie outside the productive-coupling budget.

Table S3: Optical realization used in all receiver calculations. Values refer to the central source pair; the geometry, internal gaps, and productive budget are those in Table S1.

| Quantity | Value |
|---|---|
| Common excited-manifold shift | 0.2361763525 eV |
| $(\epsilon_C, \epsilon_I, \epsilon_E)$ | $(3.054176353, 2.727976611, 2.747976611)$ eV |
| $\lambda_{EG}$; vertical photon energy | 0.768 eV; 1.979976611 eV |
| Photon-spectrum peak; full width at half maximum | 594.00 nm; 135.08 nm |
| Mean photon energy | 2.043425320 eV |
| $V_{CG}$; $V_{IG}$ at $b = 2500\,\mathrm{s}^{-1}$ | $0.590053918\,\mu$eV; $7.769344435\,\mu$eV |
| Effective vacuum transition dipole | 0.0097338744 D, chosen for $r = 125\,\mathrm{s}^{-1}$ |
| Intrinsic nonradiative rate $n$ | $41.66666667\,\mathrm{s}^{-1}$, an independent phenomenological channel |

The dipole follows from $r = |\mu|^2 \int_0^\infty E^3 \mathcal{D}_{EG}(E) dE/(3\pi\epsilon_0 \hbar^4 c^3)$, with energies expressed in joules. The intrinsic loss $n$ is an independent phenomenological channel; realizing its value through the same $E \to G$ Marcus–Jortner formula at this geometry would require coupling outside the weak-coupling regime. In energy–loss scans the optical $E, G$ surfaces stay fixed, and the loss coupling is adjusted to the chosen terminal rate. Hence $\rho_\gamma$ and the receiver spectral projection remain common across Figure 4(c).

The shifted construction gives a photon-spectrum peak at 594 nm with the full bandwidth retained. Since the internal and terminal rates are preserved, the source generator and the source-only results in Figures 1–3 are unchanged.

## SI1.4 Physical interpretation and experimental connections

In the harmonic realization, $\Delta_i$ denotes a separation of relaxed potential minima, not automatically a trap-release activation barrier. For a material assignment, restricted partition functions include the state energies and internal multiplicities in the thermal weights (SI1.2). The couplings and reorganization parameters describe interconfiguration transfer; $a$, $b$, and $n$ describe termination.

The local mean visit duration in $I$ is $(b+v+w)^{-1} \simeq 0.258$ ms at the task point. Transfer and loss both end a visit, while feeding and repeated returns determine the full population decay and emission tail.

**A representative material interpretation: ZnS:$Mn^{2+}$.** For delayed excitation, Mukhina et al. discuss donor-electron recombination with a Mn-associated hole and an alternative route through a Mn-bound exciton.[S8] This proposed scheme gives a candidate correspondence: $C$ is the prepared local carrier pair, $I$ a Mn-associated bound-exciton configuration, $E$ the intraionic $Mn^{2+}$ $^4T_1$ excitation, and $G$ the terminated configuration with Mn in $^6A_1$. Direct recombination can populate $E$ without another resolved stage ($C \to E$); formation of $I$ followed by energy transfer gives $C \to I \to E$. Both paths describe the relaxation of one local carrier pair. A population-resolved bound exciton supplies the explicit $I$ state; a rapidly eliminated mediator instead contributes to an effective transfer rate.

Spatially distinct ML-active and relatively PL-dark regions[S8] motivate preparation-resolved measurements. Comparing their normalized yields requires calibrated absorption, initial populations, and collection efficiency. Host-excited PL can populate carrier configurations before $E$, so its initial-state distribution must also be established. For the local-pair interpretation, relaxation is measured at fixed strain after preparation, with negligible continuing injection or mixing between pairs and the reduction conditions of SI1.2. The monitored Mn channel and its multiplet weights specify the optical readout and thermal factors.

**Experimental parameter constraints.** Table S4 lists prospective measurements to constrain the state assignment and parameters for one sample under fixed relaxation conditions.

Changing Mn concentration or defect density can alter preparation, transfer, and termination together, so a measured material series requires updated coefficients for the budget comparison.

**Normalization for a quantitative comparison.** When the model observables are to be estimated for a particular sample, let $n_{\gamma,i}$ be the mean background-corrected full-time detected photon count, $N_i$ the independently calibrated number of initial excitations in $i$, and $\zeta_i$ the effective collection and detection efficiency for the monitored channel. Then

$$n_{\gamma,i} = \zeta_i N_i \Phi_i, \qquad q_C = \frac{n_{\gamma,C}}{n_{\gamma,E}} \frac{\zeta_E N_E}{\zeta_C N_C}. \tag{S10}$$

The normalization requires $N_C, N_E$, their preparation fractions, and the collection efficiencies. Equal mechanical loading alone does not calibrate $N_C$. Optical absorption constrains $N_E$ when branching into the modeled initial state is known. Finite preparation duration and detector response are included when extracting emission times.

**Inference and independent closure tests.** With $r, \kappa, a, b$ and the thermal weights independently constrained as in Table S4, the measured yields predict the PL mean through Eq. (12); that mean is reserved for validation. If these inputs are inferred from the same records, the relation instead provides a conditional consistency check.

Table S4: Prospective parameter constraints for the candidate $ZnS:Mn^{2+}$ correspondence.

| Quantity | Measurement or analysis | What it can establish |
|---|---|---|
| $E, G$; $r, n$ | Spectrally resolved ML/PL, absolute PL yield, and time-resolved PL. | The monitored Mn emission identifies the radiative channel. Only when escape from $E$ is negligible does an intrinsic single-exponential lifetime $\tau_{\text{iso}}$ give $r = \Phi_{\text{iso}}/\tau_{\text{iso}}$ and $n = (1 - \Phi_{\text{iso}})/\tau_{\text{iso}}$. |
| $C, I$; transfer rates | State-sensitive transient absorption or charge-state spectroscopy, combined with emission following independently characterized preparations. | Jointly assigned population and emission records constrain distinguishable transfer branches; individual rates require identifiable, nonredundant measurements.[S9] |
| $a, b$ | Joint analysis of nonemissive-state populations, return/transfer branches, and calibrated photon yields. | The local escape rate, combined with independently constrained $v, w$, determines $b$; fed or returning populations require the full kinetic analysis. Equation (14) provides an alternative inference. |
| $\chi_I$, $\Delta_I$ | Forward/reverse rate ratios at fixed temperature, supplemented by state and multiplet spectroscopy. | For nonzero $w, z$ satisfying detailed balance, $\chi_I = z/w$ gives $F_I^{\text{th}} - F_E^{\text{th}} = -k_B T_b \ln(z/w)$. |
| $N_C$, $\Phi_C$ | Absolute photon collection together with an independently calibrated initial local-pair population. | Calibrated $N_C$ and collection efficiency give the per-preparation yield through Eq. (S10). |

Alternatively, Eq. (14) estimates an unknown $b$ from the yields and mean. The numerator and squared denominator must be positive and resolved relative to their uncertainties, and the probability constraints in Eq. (12) must hold. Both procedures retain covariances from shared photon records. Validation of the inferred $b$ uses a separately constrained loss measurement or additional records reserved from the fit.

# SI2 Receiver parameters, normalization, and periodic propagation

The four-state receiver has two closed states, $C_1, C_2$, and two open states, $O_1, O_2$, with $C_1(0) = 1$:

$$\begin{aligned}
\dot{C}_1 &= -G_{a1}C_1 + G_{d1}O_1 + G_rC_2, \\
\dot{O}_1 &= G_{a1}C_1 - (G_{d1} + G_f)O_1 + G_bO_2, \\
\dot{O}_2 &= G_fO_1 - (G_b + G_{d2})O_2 + G_{a2}C_2, \\
\dot{C}_2 &= G_{d2}O_2 - (G_{a2} + G_r)C_2.
\end{aligned} \tag{S11}$$

The relative conductance is $g = O_1 + 0.05O_2$. With the channel-weighted normalized fluxes $x_j$ of Eq. (29),

$$G_{ai} = k_i \frac{x_{ai}}{1 + x_{ai}}, \qquad G_f = G_{f0} + k_f \frac{x_f}{1 + x_f}, \qquad G_b = G_{b0} + k_b \frac{x_b}{1 + x_b}. \tag{S12}$$

Equations (31), (S11), and (S12) define the receiver dynamics. The reduced model uses the ChrimsonR closing-time anchor $\gamma^{-1} = 15.8$ ms,[S10] with $\beta = 0.2$ and $k_r = 10\,\mathrm{s}^{-1}$ specified separately. The four-state equations and rates in Table S5 use the model and parameter table of Gupta et al.,[S11] whose principal dynamical calibration concerns vf-Chrimson. Their table also lists closing rates for Chrimson and f-Chrimson. We retain the shared parameters, with both flux exponents equal to one, and select the Chrimson entry $G_{d1} = 41\,\mathrm{s}^{-1}$ ($0.041\,\mathrm{ms}^{-1}$). These shared rates with the selected closing rate define the Chrimson-based benchmark. The reduced model uses the ChrimsonR closing-time anchor, whereas the optical projection below uses the Chrimson–EYFP initial-response spectrum.

Table S5: Four-state reference-model parameters and exposure conventions.

| Quantity | Value or definition |
|---|---|
| $G_{d1}$, $G_{d2}$, $G_r$ | 41, 10, 0.000667 $\mathrm{s}^{-1}$ |
| $k_1$, $k_2$ | 3000, 200 $\mathrm{s}^{-1}$ |
| $G_{f0}$, $G_{b0}$ | 20, 3.2 $\mathrm{s}^{-1}$ |
| $k_f$, $k_b$ | 10, 10 $\mathrm{s}^{-1}$ |
| Measured model response | $g = O_1 + 0.05O_2$ |
| Reference wavelength and flux | 594 nm; $\phi_m = 1.5 \times 10^{16}$ photons $\mathrm{mm}^{-2}$ $\mathrm{s}^{-1}$ |
| Common preparation | $x_j = \xi_j q I_C/(k_1\Phi_{C0})$; incident normalized fluence $\mathcal{F}_{\mathrm{inc}} = (q/k_1)(\Phi_C/\Phi_{C0})$ |
| Equal incident photon fluence | $x_j = \xi_j q I_C/(k_1\Phi_C)$; $\mathcal{F}_{\mathrm{inc}} = q/k_1$ |
| Isolated initial state | $(C_1, O_1, O_2, C_2) = (1, 0, 0, 0)$ |
| Repeated stimulation | Period $T = 50$ ms; per-pulse $q$ held fixed |

The incident normalized fluence $\mathcal{F}_{\mathrm{inc}}$ has units of time and is defined before action-spectrum weighting. Its photon fluence is $N_\gamma = \phi_m \mathcal{F}_{\mathrm{inc}}$, whereas $\int x_j dt = \xi_j \mathcal{F}_{\mathrm{inc}}$ is a channel-weighted quantity. Incident optical energy is $E_{\mathrm{inc}} = \langle E_\gamma \rangle \phi_m \mathcal{F}_{\mathrm{inc}}$, with the mean photon energy $\langle E_\gamma \rangle = 2.04343$ eV for the calculated spectrum.

## SI2.1 Experimental spectral proxy and broadband receiver input

We use the Chrimson-only series from Figure 1D source data of Govorunova et al.[S12] The construct is Chrimson–EYFP. The published action spectrum uses initial photocurrent slopes from low-intensity 15 ms pulses, corrected for photon density. The archived workbook contains 30 wavelengths from 400 to 690 nm, with means and SEMs from four cells. We take the magnitude of the inward-current series and normalize its interpolation to its value at 594 nm. The measured Chrimson–EYFP action spectrum includes the EYFP-associated shoulder near 520 nm.

The baseline uses linear interpolation within the measured range. Outside it, positive exponential tails have a blue decay length of 40 nm and a red decay length of 15.5812 nm, the latter fitted to the logarithms of the last five mean values. The source emits 88.81% of its photons within the measured interval. Integrating the whole spectrum gives $\xi = 0.61133627$ [Figure S1(a)]. A spectrally varying optical path can be included by multiplying the integrand in Eq. (29) by its relative transmission; the baseline assumes this transmission is unity, with the common overall coupling already in $q$.

The initial-response spectrum supplies a proxy for $\mathcal{A}_{a1}$. The baseline approximates the other channel factors by $\xi_{a2} = \xi_f = \xi_b = \xi_{a1}$ and applies the empirical 594 nm saturation law to the wavelength-integrated drives in Eq. (S12). This broadband constitutive approximation sums photons before saturation. The reduced receiver shares the first-channel projection as a normalization control.

## SI2.2 Monochromatic control and limits of the shared-factor approximation

For a common, time-independent spectral factor $\xi$, every light-dependent rate depends on $q$ through $\xi q$. This gives, for either receiver and any fixed preparation protocol,

$$y_{\text{broad}}(t; q) = y_{594}(t; \xi q), \qquad q_{\text{min,broad}} = q_{\text{min},594}/\xi, \tag{S13}$$

where the cost relation requires the scaled feasible minimum to remain in the search domain. The same common scaling applies to finite preparation and periodic inputs. Channel-dependent factors and evolving spectra require the individual drives of Eq. (29); source pairs with different common factors have separate exposure rescalings.

Table S6: Monochromatic and broadband responses at common preparation $q = 1$. Means use $W = 20$ ms. The final column gives the common preparation-cost multiplier relative to monochromatic input for the activation–closure task; absolute costs are listed in Table S7.

| Receiver | Input | Direct mean | Relay mean | Cost multiplier |
|---|---|---|---|---|
| Three-state | 594 nm single line | 0.30110 | 0.34501 | 1 |
| Three-state | Broad spectrum | 0.20656 | 0.24574 | 1.63576 |
| Four-state | 594 nm single line | 0.29939 | 0.32307 | 1 |
| Four-state | Broad spectrum | 0.20740 | 0.23577 | 1.63576 |

Broadband weighting increases both preparation requirements by $1/\xi = 1.63576$. Equation (S13) preserves their ratio and hence the 16.54% four-state saving. At fixed $q = 1$, absolute responses decrease; the relative four-state gain increases from 7.91% to 13.68% because the comparison moves along a nonlinear exposure–response relation.

For a prescribed channel-factor stress test, we independently set $(\xi_{a2}, \xi_f, \xi_b)/\xi_{a1}$ to every combination of $\{0.5, 1, 2\}$, holding $\xi_{a1} = 0.61133627$ and all dark rates fixed. Across these 27 specified stress tests, both sources meet the central task and the saving ranges from 16.532% to 16.544% [Figure S1(b)]. At matched activation in the baseline, the integrated $C_2 \to O_2$ event count is only $6.47 \times 10^{-5}$ and $1.71 \times 10^{-6}$ of the corresponding $C_1 \to O_1$ count for direct and relay inputs; $O_2$ contributes 0.79% and 1.03% of integrated conductance. First-channel dominance explains the weak sensitivity in this brief, dark-initialized task. Applying the benchmark to adapted or sustained stimulation requires history-dependent channel spectra.[S13]

Spectral controls give $\xi = 0.61084$–$0.61279$ when the unmeasured tails are respectively set to zero or held at their endpoint values, and 0.61233 with shape-preserving cubic interpolation. The bracket assumes nonincreasing unmeasured tails without additional absorption bands.

A uniform preparation-loading profile uses $s_p(t) = 1/t_p$ for $0 \leq t \leq t_p$ and zero otherwise. Its convolution is evaluated analytically for each exponential source mode. Finite trains sum emissions from all preceding preparations beginning at time zero. The periodic optical input instead uses Eq. (S21), with the geometric tail sum evaluated analytically. These inputs are propagated separately from receiver initialization.

For a prescribed periodic optical waveform, either receiver evolves linearly in its population vector even though the rates depend nonlinearly on the incident flux. Propagating a basis over one period gives a stochastic period map $M_T$. The periodic population is its normalized fixed point, $M_T \boldsymbol{p}_* = \boldsymbol{p}_*$. Integrated response is propagated as an additional accumulator. Finite records start from the dark state. Whole-train and last-period means use Eqs. (37) and (S20), respectively.

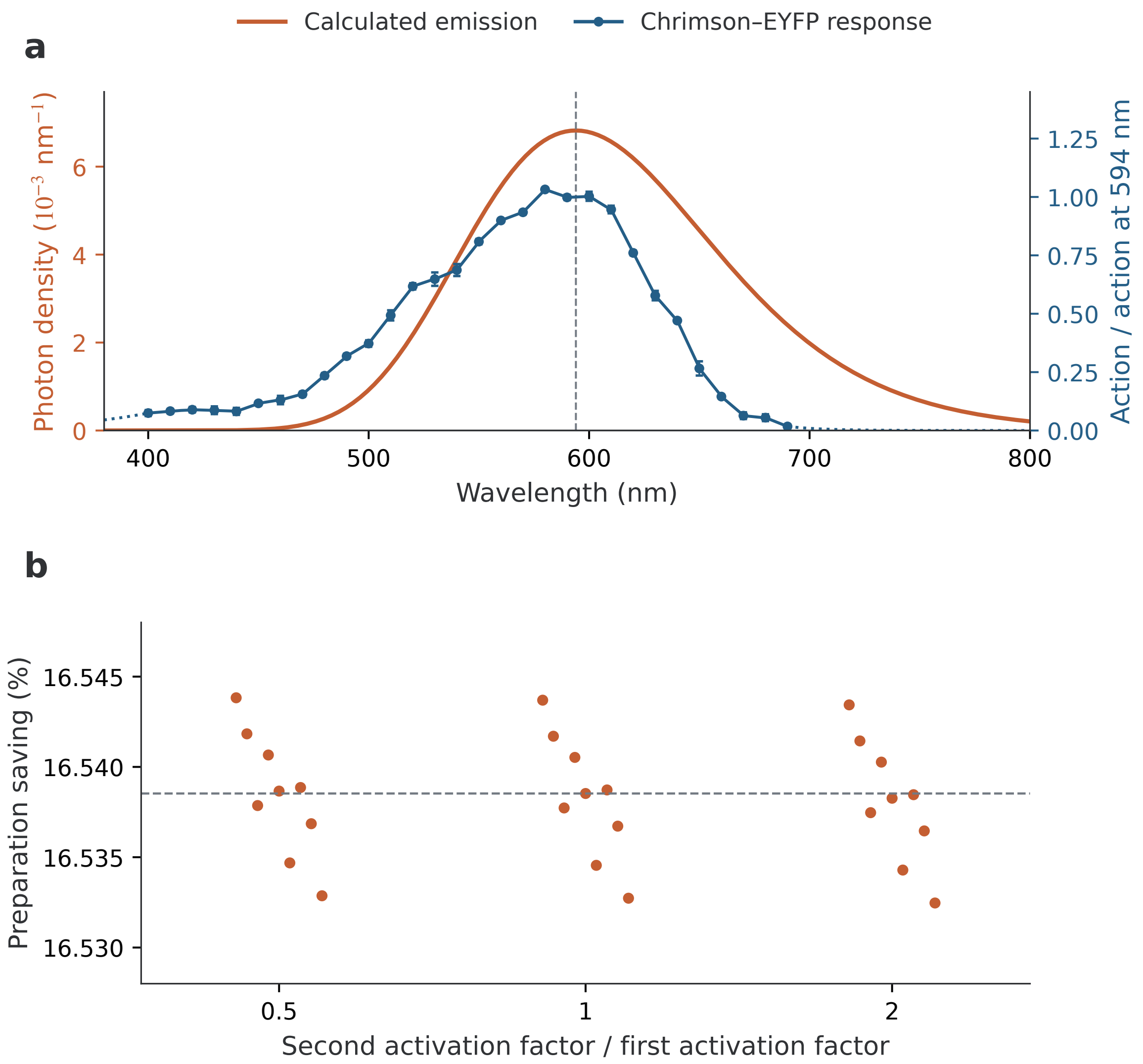

Figure S1: Explicit spectral interface and a test beyond common rescaling. (a) Calculated photon density (orange, left axis) and the Chrimson–EYFP initial-response action spectrum normalized at 594 nm (blue, right axis). Points and error bars show the published means and SEMs divided by the fixed baseline 594 nm normalizer. Solid blue segments interpolate measured wavelengths; dotted tails are assumptions. The vertical dashed line marks the photon-spectrum peak. (b) Four-state preparation saving in the 27 independent channel-factor tests, grouped by $\xi_{a2}/\xi_{a1}$. Each point represents one combination of $\xi_f/\xi_{a1}$ and $\xi_b/\xi_{a1}$; small horizontal offsets distinguish them. The dashed line is the shared-factor result. All tests use the central activation–closure task and fixed source pair.

# SI3 Allocation boundary limits and finite-window collection

This section gives the boundary cases of the complete-yield endpoint theorem, the short-time collection expansion, and a counterexample showing that the theorem does not extend to hard-cutoff collection for all reversible rate networks.

## SI3.1 Complete-yield boundary conventions

The endpoint theorem in Section 4 assumes $a, c_0, r > 0$. Its zero-link cases follow from the same integrated elimination as Eq. (19). At $s = 0$, $u = v = 0$, so $c_{\text{eff}} = c \leq c_0$, $a_{\text{eff}} = a$, and $\kappa_{\text{eff}} = \kappa + \chi_I wb/(b + w) \geq \kappa$. At $s = 1$, $w = z = 0$, so $c_{\text{eff}} = c \leq c_0$, $\kappa_{\text{eff}} = \kappa$, and $a_{\text{eff}} = a + ub/(b + v) \geq a$. In either case the yield cannot exceed the direct reference. A vanishing denominator in these expressions corresponds to a decoupled intermediate, which is omitted.

For $b = 0$, $p, h > 0$, and $\eta > 0$, elimination instead gives

$$c_{\text{eff}} = c_0(1 - \eta) + \frac{ph}{\nu + h}\eta, \qquad a_{\text{eff}} = a, \qquad \kappa_{\text{eff}} = \kappa. \tag{S14}$$

The yield is an increasing function of $c_{\text{eff}}$, so this affine interpolation is bounded by its endpoints. Equality along the whole ray is possible when $ph/(\nu + h) = c_0$. Unreachable states are omitted before matrix inversion; if a preparation cannot reach $E$, its photon yield is zero and its photon-conditioned mean is undefined.

## SI3.2 Short-time photon collection

For sufficiently short windows, direct transfer must win. Expanding Eq. (28) gives

$$F_W = \frac{rc}{2}W^2 + \frac{r}{6}\{uw - c(a + c + u + \kappa + d + z)\}W^3 + O(W^4). \tag{S15}$$

The direct route contributes at order $W^2$, while a pure relay requires two transfers and starts at order $W^3$. For bounded coefficients on the compact budget domain, the mixed-minus-direct difference is $-rc_0\eta W^2/2 + \eta O(W^3)$ uniformly. Direct transfer therefore maximizes collection for sufficiently small $W$; any relay advantage begins at a finite window length.

## SI3.3 A finite-window counterexample

A mixed allocation can outperform both direct transfer and every pure relay before a finite deadline. Retain $a, r, \kappa, c_0, \chi_C$ from Table S1, but choose $u_B = 3.0 \times 10^6\,\text{s}^{-1}$, $w_B = 1.0 \times 10^5\,\text{s}^{-1}$, $b = 4.5 \times 10^5\,\text{s}^{-1}$, and $\chi_I = 20$. At $W = 0.12$ ms, the fixed mixed allocation $(\eta, s) = (0.043, 0.42)$ gives $F_W = 1.41542099970 \times 10^{-5}$, whereas the pure-direct value is $1.36004560098 \times 10^{-5}$.

For $0 < s < 1$, the pure-relay total yield is

$$\Phi_C(1, s) = \frac{r}{\kappa + \chi_C a + \chi_I b + A_*/s + B_*/(1 - s)}, \tag{S16}$$

$$A_* = \frac{a(\kappa b/w_B + \kappa + \chi_I b)}{u_B}, \qquad B_* = \frac{\kappa(ab/u_B + a\chi_C/\chi_I + b)}{w_B}. \tag{S17}$$

Cauchy–Schwarz, or direct minimization, gives $A_*/s + B_*/(1 - s) \geq (\sqrt{A_*} + \sqrt{B_*})^2$, with equality at $s = \sqrt{A_*}/(\sqrt{A_*} + \sqrt{B_*})$. Since finite-window emission cannot exceed total emission,

$$\max_s F_W(1, s) \leq \frac{r}{\kappa + \chi_C a + \chi_I b + (\sqrt{A_*} + \sqrt{B_*})^2} = 1.38813049182 \times 10^{-5}. \tag{S18}$$

The mixed design exceeds this upper bound by 1.9660%. These coefficients define a separate reversible rate family with the same allocation rule as Eq. (4). Its total yield, $1.43767558352 \times 10^{-5}$, remains below the direct value, consistent with the complete-yield endpoint theorem.

The collection protocol itself matters. Exponentially weighted output $\int_0^\infty e^{-\lambda t} I_C(t)dt$ is exactly the total yield of the generator $L - \lambda\mathbb{1}$, with $a, b, \kappa$ increased by $\lambda$ and $r$ unchanged. Detailed balance and the allocation structure survive, so the total-yield endpoint theorem applies. A hard cutoff is not equivalent to that weighting.

# SI4 Task search and numerical feasibility

The central task has $W = 20$ ms, $y_* = 0.20$, and $T_c = 50$ ms. Exposure searches cover $10^{-5} \leq q \leq 10^3$, initially on 241 logarithmic points. Roots of activation and closure boundaries are resolved, and the resulting intervals are examined for feasible minima. A separate check samples 961 points above the activation threshold and refines local residual minima. A missing feasible solution refers only to this searched domain and numerical resolution. This interval search includes nonmonotone closure responses.

Table S7: Fixed-source task costs at the common residual allowance $\epsilon_c = 0.05$.

| Receiver | Source | $q$ | $y(T_c)$ | Saving |
|---|---|---|---|---|
| Three-state | Direct | 0.961191804 | 0.035229597 | — |
| Three-state | Relay | 0.767574984 | 0.016299230 | 20.1434% |
| Four-state | Direct | 0.955562259 | 0.046531210 | — |
| Four-state | Relay | 0.797526324 | 0.024818409 | 16.5385% |

The activation-boundary means equal 0.20 to numerical precision. The matched traces in Figure 5(b) use an exposure margin of $1 + 10^{-7}$ and independent tight propagation to ensure feasibility. Their four-state exposures are $q = 0.95556235$ and $0.79752640$, with residuals 0.04653121 and 0.02481841 for direct and relay inputs, respectively. These feasible trajectories provide numerical upper bounds on the minimum costs.

Table S8 gives the four-state feasible thresholds. The reduced receiver additionally permits the fixed relay to satisfy $\epsilon_c = 0.02$.

Table S8: Four-state searched minimum preparation at selected residual limits, with $W = 20$ ms, $y_* = 0.20$, and $T_c = 50$ ms. "Not found" refers to the search specified above.

| $\epsilon_c$ | Direct $q$ | Fixed relay $q$ |
|---|---|---|
| 0.01 | Not found | Not found |
| 0.02 | Not found | Not found |
| 0.03 | Not found | 0.79753 |
| 0.05 | 0.95556 | 0.79753 |
| 0.10 | 0.95556 | 0.79753 |

# SI5 Numerical methods

Source rates are evaluated with positive thermal sideband sums and detailed-balance reverse rates. Yields and moments use linear solves on the accessible transient subspace. A reversible similarity transform permits symmetric diagonalization, and source modal integrals are evaluated analytically without a finite time cutoff. Small-time exponential differences avoid cancellation in the initial zero-current limit.

Receiver populations and the integrated response are propagated together using DOP853 with relative tolerance $2 \times 10^{-9}$ and absolute tolerance $2 \times 10^{-12}$; stiff high-exposure reduced-model calculations use LSODA. Integrations split at preparation-pulse and repetition boundaries. Independent Radau propagation and simultaneous source–receiver integration check representative isolated, finite-train, and periodic cases.

# SI6 Photon-window, normalization, and history controls

## SI6.1 Collection windows and receiver closing

Figure S2 distinguishes the finite-window photon ratio, finite-window conductance ratio, and complete-yield ratio for the fixed source pair. The early photon gain is attenuated in the receiver response.

Increasing $G_{d1}$ alone while keeping the other four-state rates fixed further changes the ranking. At $G_{d1} = 175$ and $370\,\mathrm{s}^{-1}$, the response ratios at $q = 1$, $W = 20$ ms are approximately 0.950 and 0.871. These two rates are drawn from the faster-variant entries in Gupta et al.;[S11] the scan varies only $G_{d1}$, keeping the other rates and the spectrum fixed.

## SI6.2 Common preparation and equal delivered fluence

Common preparation uses $x_j = \xi_j q I_C/(k_1 \Phi_{C0})$ and preserves the relay's lower complete yield. To isolate waveform effects at equal incident photon fluence, we instead use

$$x_j(t) = \frac{\xi_j q}{k_1}\frac{I_C(t)}{\Phi_C}, \qquad \mathcal{F}_{\text{inc}} = \frac{q}{k_1}, \quad \int_0^\infty x_j(t)\,dt = \frac{\xi_j q}{k_1}. \tag{S19}$$

Equal incident fluence generally requires different prepared populations. Figure S3(a) shows the resulting four-state map; panel (b) returns to common preparation for the three-state receiver. The regions differ quantitatively because both the normalization and the receiver kinetics affect the comparison.

## SI6.3 Source startup, receiver history, and averaging

A last-period mean,

$$\overline{y}_N = \frac{1}{T}\int_{(N-1)T}^{NT} y(t)\,dt, \tag{S20}$$

differs from the whole-train mean in Eq. (37). At $q = 1$, the four-state relay/direct ratio in the twentieth period is about 1.035, although its twenty-period whole-train ratio is 0.953. Early periods remain in the latter average. The separately computed periodic limit has ratio 0.882.

For a source current $I_C(t) = \sum_j \widetilde{A}_j^{(C)} e^{-\lambda_j t}$, the rate amplitudes $\widetilde{A}_j^{(C)}$ need not be nonnegative individually, although the total current is nonnegative. An already-periodic input is

$$I_{C,T}(t) = \sum_j \frac{\widetilde{A}_j^{(C)} e^{-\lambda_j t}}{1 - e^{-\lambda_j T}}, \qquad 0 \le t < T. \tag{S21}$$

Figure S4 compares this waveform, applied to a dark receiver, with a finite source train beginning at the first preparation. The nearly coincident last-period ratios show that the transient ranking change persists with a periodic source input, identifying a separate contribution from receiver-state accumulation.

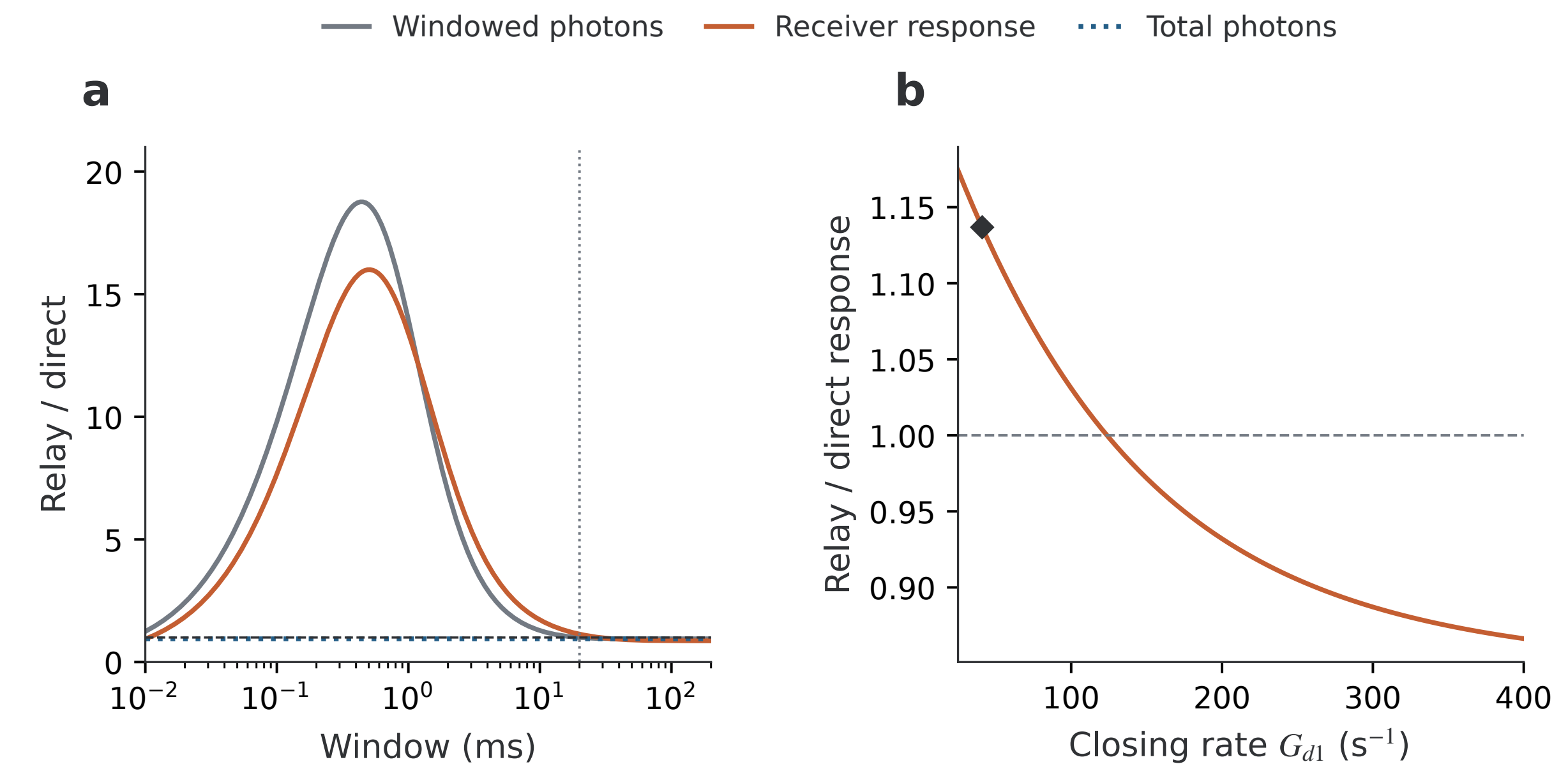

Figure S2: Additional temporal controls for the fixed source pair. (a) Ratios of photons collected before a deadline and integrated four-state conductance at $q = 1$, together with the complete-yield ratio. The vertical dotted line marks 20 ms. (b) Four-state response ratio at $q = 1$, $W = 20$ ms when only the closing rate $G_{d1}$ varies; the diamond marks the reference value $41\,\mathrm{s}^{-1}$. Horizontal dashed lines denote equal responses or photon counts.

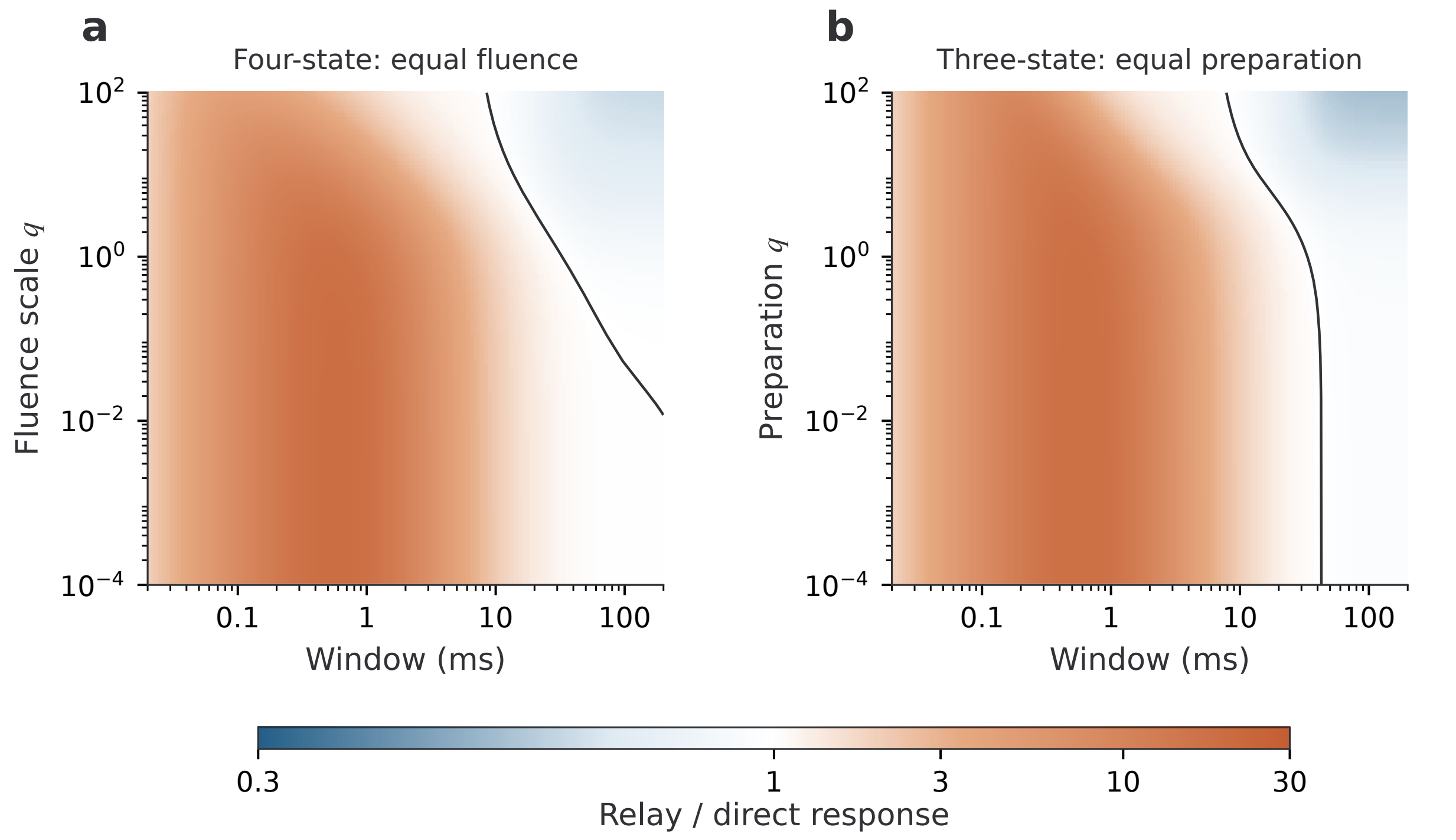

Figure S3: Normalization and reduced-receiver controls. Relay/direct integrated-response ratios for (a) the four-state receiver at equal normalized incident photon fluence $q/k_1$ and (b) the three-state receiver at common preparation $q$. The fixed source pair is used throughout. Black contours denote unity; the common color scale is logarithmic on each side of unity. The parameter $q$ in (a) sets delivered fluence rather than common preparation.

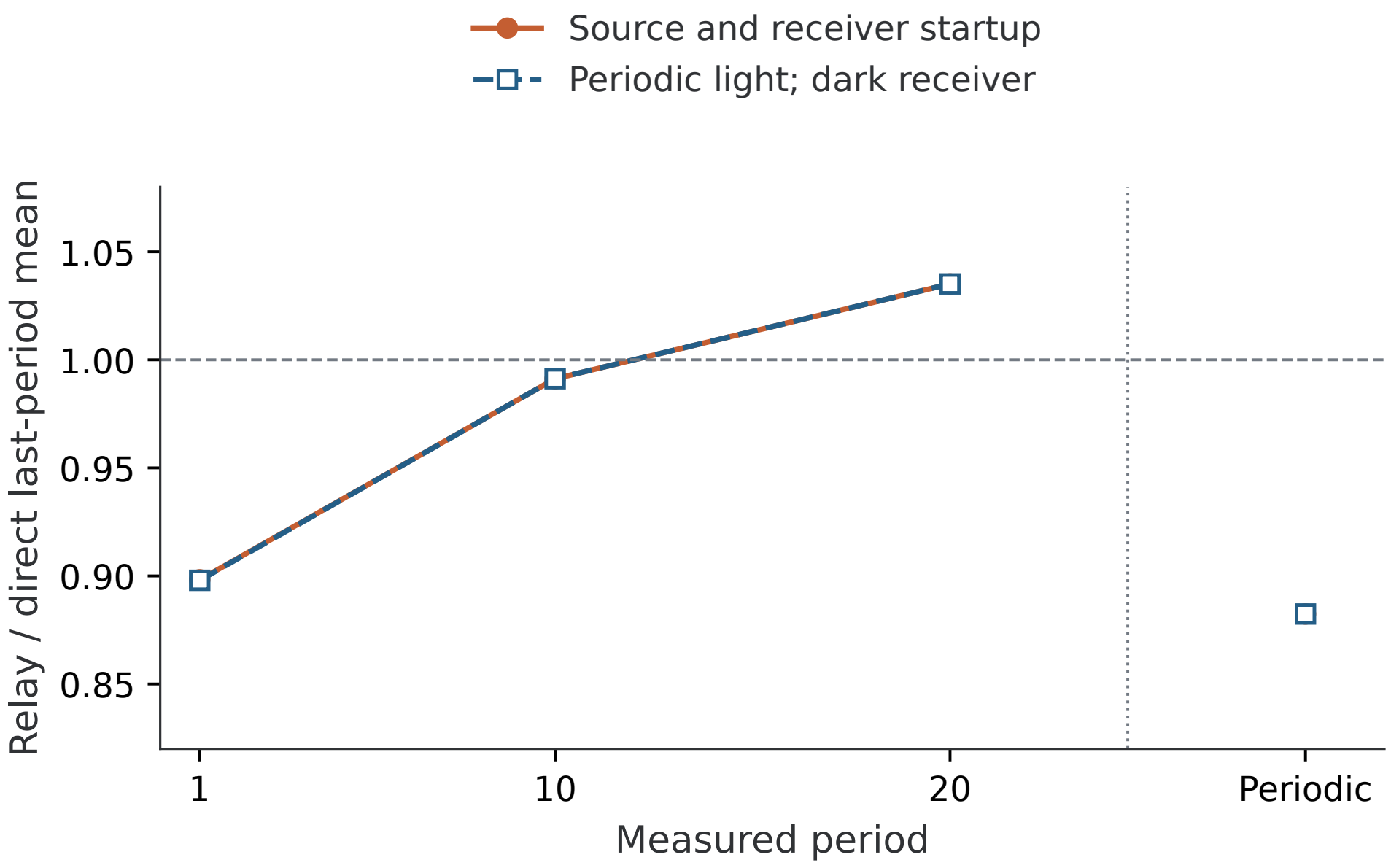

Figure S4: History control for the four-state receiver at $q = 1$ and $T = 50$ ms. Last-period mean-conductance ratios are compared when both source and receiver start from rest and when an already-periodic optical input is applied to the same dark receiver. Finite-period points use their actual period numbers; periodic stationary values are shown separately. These last-period means differ from the whole-train means in Figure 6(c).